\documentclass[journal=ancac3,manuscript=article,layout=traditional]{achemso}
\setkeys{acs}{articletitle=true}
\usepackage[utf8]{inputenc}

\usepackage[version=3]{mhchem} 
\usepackage[T1]{fontenc}       
\usepackage{epsfig}
\usepackage{subcaption}
\usepackage{natbib}
\usepackage{caption}
\usepackage{float}
\usepackage{subfloat}
\usepackage{geometry}
\usepackage{setspace}
\usepackage{xkeyval}
\usepackage{amsmath}
\usepackage{amssymb} 
\usepackage{newtxtext}
\usepackage{color}
\usepackage{longtable}
\usepackage{makecell}

\usepackage{bm}

\usepackage{siunitx} 
\DeclareSIUnit\molar{\mole\per\cubic\deci\metre}
\DeclareSIUnit\Molar{M}

\usepackage{amsfonts}
\usepackage{dsfont}
\usepackage[space]{grffile}
\author{Clemens Jochum} 
\affiliation{Institute for Theoretical Physics, TU Wien, Wiedner Hauptstra{\ss}e 8-10, A-1040 Vienna, Austria} 

\author{Nata\v{s}a Ad\v{z}i\'{c}}
\affiliation{Faculty of Physics, University of Vienna, Boltzmanngasse 5, A-1090 Vienna, Austria}
\email{natasa.adzic@univie.ac.at}

\author{Emmanuel Stiakakis} 
\affiliation{Institute of Complex Systems 3, Forschungszentrum J\"ulich, Leo-Brandt-Stra{\ss}e, D-52425 J\"ulich, Germany}

\author{Thomas L. Derrien}
\affiliation{Department of Biological and Environmental Engineering, Cornell University, Ithaca, New York 14853-5701, USA}

\author{Dan Luo}
\affiliation{Department of Biological and Environmental Engineering, Cornell University, Ithaca, New York 14853-5701, USA}

\author{Gerhard Kahl} 
\affiliation{Institute for Theoretical Physics, TU Wien, Wiedner Hauptstra{\ss}e 8-10, A-1040 Vienna, Austria}
\alsoaffiliation{Center for Computational Materials Science (CMS), TU Wien, Wiedner Hauptstra{\ss}e 8-10, A-1040 Vienna, Austria}

\author{Christos~N.~Likos} 
\affiliation{Faculty of Physics, University of Vienna, Boltzmanngasse 5, A-1090 Vienna, Austria}

\title{Structure and stimuli-responsiveness of all-DNA dendrimers: theory and experiment}

\begin{document}


\begin{tocentry}
\includegraphics[width=6cm]{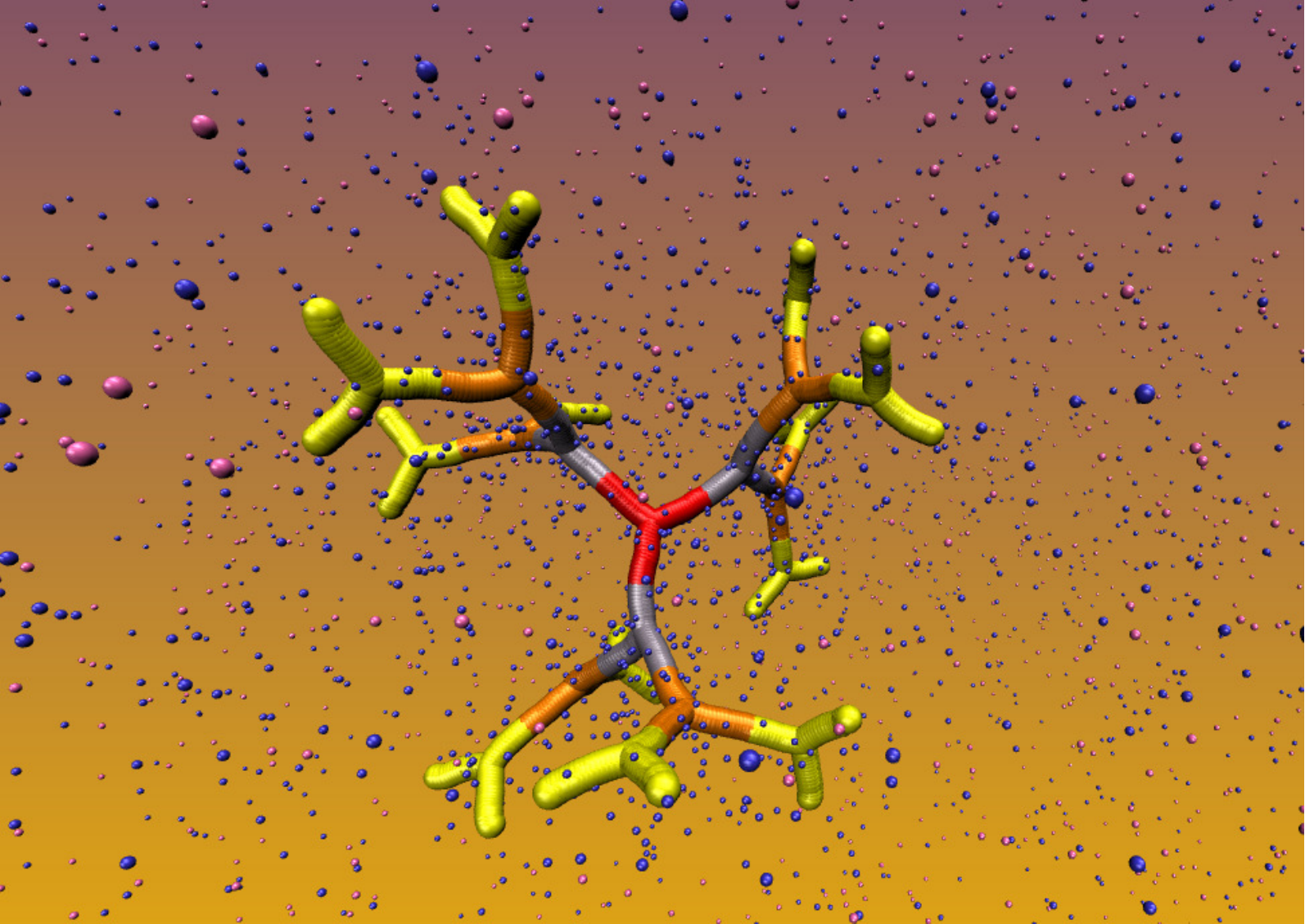}



\end{tocentry}

\begin{abstract} 

We present a comprehensive theoretical and experimental study of the solution phase properties of DNA-based family of nanoparticles - dendrimer-like DNA molecules (DL-DNA).  
These charged DNA dendrimers are novel macromolecular aggregates, which hold high promise in  targeted self-assembly of soft matter systems in the bulk and at interfaces.
To descibe the behavior of this family of dendrimers (with generations ranging from G1 to G7), we use a theoretical model in which base-pairs of a single DL-DNA molecule 
are modeled by charged monomers, whose interactions are chosen to mimic the equilibrium properties of DNA correctly.
Experimental results on the sizes and conformations of DL-DNA are based on static and dynamic light scattering; at the same time,
Molecular Dynamics simulations are employed to model the equilibrium properties of DL-DNA, which compare favorably with the findings
from experiments while at the same time providing a host of additional information and insight into the molecular structure of the nanostructures.
We also examine the salt-responsiveness of these macromolecules, finding that despite the strong screening of electrostatic interactions, 
brought about by the added salt, the macromolecules shrink only slightly, their size robustness stemming from the high bending rigidity of the DNA-segments.  
The study of these charged dendrimer systems is an important field of research in the area of soft matter due to their potential role for various interdisciplinary applications, 
ranging from molecular cages and carriers for drug delivery in a living organism to the development of dendrimer- and dendron-based ultra-thin films in the area of nanotechnology. These findings are essential to determine if DL-DNA is a viable candidate for the experimental realization of cluster crystals in the bulk, a novel form of solids with multiple site occupancy.

\end{abstract}

\\
\\

All-DNA constructs are complex self-assemblies made solely by DNA. The creation of such nanostructures was initiated in the early 1980s when Seeman proposed the use of DNA as a 
programmable nanoscale building material~\cite{a1}, laying the foundation  for structural DNA nanotechnology~\cite{NatNanotech2011,NatRevMat2017}. This interdisciplinary research field has had a striking 
impact on nanoscience and nanotechnology, demonstrating the construction of a remarkably rich assortment of multidimensional all-DNA 
nanoarchitectures~\cite{Nature1983,Biochemistry1993,Nature1998,Nature2008,Nature2006,Nature2009,Science2012} with promising applications in areas such as molecular and 
cellular biophysics~\cite{PNAS2007,Jacks2010,AngewChem2009,NatureMeth2014,Jacks2014}, macromolecular crystallography~\cite{Nature2009p1}, inorganic nanoparticle templated self-assembly for 
nanoelectronics~\cite{Nature2012,Science2012p2, ChemRev2017}, protein assembly~\cite{PNAS2007,NatNanot2014,NanoLett2011,Science2012p3,Science2017}, drug delivery~\cite{Nature2009p2} and 
biotechnology~\cite{Nature2017}. 

Very recently, the area of DNA-based self-assembly has been embraced by the research field of soft-matter physics for fabricating all-DNA particles with engineered shape and interaction potentials that could 
serve as model systems for exploring unconventional bulk phase behaviour of diverse  states of matter such as gels~\cite{PRL2015,PNAS2013,NatCom2016} 
and liquid crystals~\cite{NatCom2016p1,NatMater2017}.  In 2004, Luo and co-workers~\cite{Luo}, demonstrated the fabrication of a novel dendrimer-like DNA (DL-DNA) construct. The DL-DNA particles were 
synthesized in a controlled step-wise fashion from the enzymatic ligation of Y-shaped DNA (Y-DNA) building blocks with rigid arms and specifically designed hybridization region known as "stiky-ends", leading to the formation of a highly 
charged and void-containing macromolecular assembly with tree-like architecture (see \textbf{Methods} for more details). Here, we perform a joint experimental/theoretical analysis of the shapes, sizes and form of 
these constructs, demonstrating their unique properties. We show that they are different from  both sterically and charged-dominated dendrimers, they possess a regular spherical shape with voids, and are robust to the influence of added salt. 

DL-DNA molecules are a clear example of novel functional nanostructures that can be assembled with remarkable control and subnanometer precision through programmable sticky-end cohesions. Unlike other 
chemical dendrimers, their built-in modularity allows tailored reshaping  of the dendritic scaffold in terms of surface functionalization~\cite{AccChemRes2014} and 
internal structure modification\cite{AngewChem2012},  by employing standard tools from biofunctional chemistry. Thus, these
DNA-based dendritic architectures have been envisioned to play a promising role in developing nano-barcode~\cite{NatBiotech2005,NatProtoc2006}   DNA-based vaccine~\cite{MethMolBiol2014} 
technologies, and functioning as a structural scaffold as well as a structural probe involving multiplexed molecular sensing processes~\cite{NatMater2006,ACSNano2014}. Furthermore, from the  fundamental research 
perspective, their polyelectrolyte character and inherently open architecture near their center of mass, endow the DL-DNA particles  with an ultrasoft repulsive potential~\cite{likos2006} and penetrability; features which make the DL-
DNA molecules optimal candidates for the experimental realization of recently proposed cluster-crystal structure~\cite{clusterc1,clusterc2,clusterc3}. 

The investigation of the structural properties of DL-DNA's at a single particle level, 
including their responsiveness to charge screening, is imperative for the development of emerging applications and for the 
understanding of intriguing collective phenomena related to this type of novel soft materials. Broadly speaking, access on global molecular characteristics such as particle diameter and spatial structure at a very 
coarse-grained level, is feasible with numerous scattering techniques. However, probing in detail the internal and surface morphology of a complex nanostructure is a challenging task. 
To this end, we adopted an approach of tackling the above issues by combining experiments and simulations to profound insights from the latter into quantities
and properties not accessible to experimental techniques. As a prerequisite for establishing the reliability of the latter, we first validate the model by comparing results from the 
simulation approach with accessible experimental findings.

\begin{figure}[h]
  \begin{center}
 \includegraphics[width=0.3\textwidth]{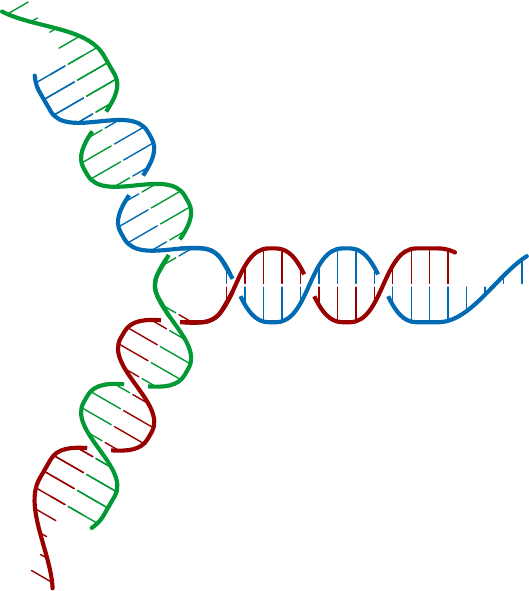}
  \end{center}
  \caption{Sketch of the Y-DNA structure: three ssDNA chains (colored red, green, and blue) assembly to form a star-like configuration with sticky ends.}
  \label{fig:fig-1}
\end{figure}

\begin{figure}[h]
  \begin{center}
 \includegraphics[width=0.8\textwidth]{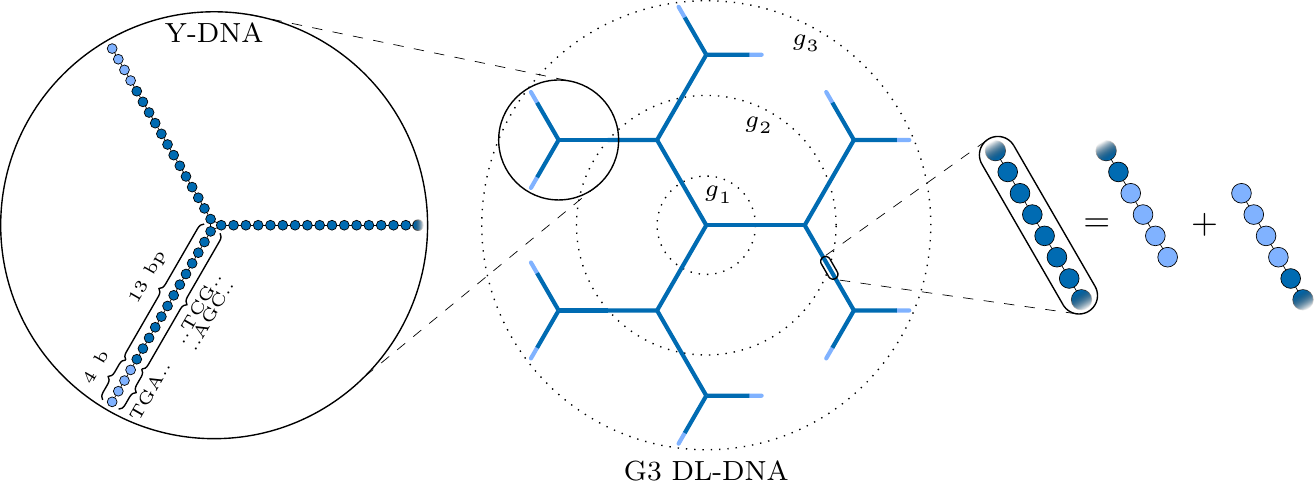}
  \end{center}
  \caption{Schematic representation of our DL-DNA model:
    the left part shows the representation of a single Y-DNA according to our particle-based model;
    the middle part shows how the union of Y-DNAs via enzymatic ligation gives rise to a dendrimer-like structure;
    the rightmost part shows the equivalence of two combined sticky ends to a regular dsDNA at the connection of two Y-DNAs in our model.}
  \label{fig:fig0}
\end{figure}

{\bf System description.} --  The building block of the dendrimer of interest is the Y-DNA unit, a three-armed structure consisting of double-stranded DNA (ds-DNA),
formed via hybridization of three single-stranded DNA chains (ss-DNA), each of which has partially complementary sequences to the other two. 
Each arm is made up of 13 base-pairs and a single-stranded sticky end with four nucleobases, see Fig.~\ref{fig:fig-1}.
While a single Y-DNA corresponds to the first dendrimer generation, G1, attaching further Y-DNA elements yields DL-DNA  of higher generations, as shown in Fig.~\ref{fig:fig0}.
This attachment is achieved by enzymatic ligation, where the single-stranded ends of two different arms form a regular double-strand through base-pairing.
In this paper we study experimentally and computationally DL-DNA macromolecules from the first generation, G1, up to the sixth generation, G6. 
We extend the theoretical model also to G7 dendrimer for selective quantities that spread light on the dendrimers' internal structure.
Subgenerations of a G$N$ dendrimer will be indicated by $\mathrm{g}_i$, $i = 1, 2, ... , N$. 

In order to build a model of the DL-DNA nanostructures, we start with a simple and widely used approach by assuming a bead-spring DNA model, where the interactions and the corresponding  parameters 
are carefully chosen to mimic the structural properties of a single ds-DNA chain~\cite{DNAm1,DNAm2}.
This particle-based model for a dsDNA, which is presented in the following, has already been introduced in recent theoretical studies of polyelectrolyte brushes \cite{DNAm1,DNAm2,ssds} and its validity has been 
tested in comparison with experiments in the context of salt-dependent forces between DNA-grafted colloids~\cite{kegler1,kegler2,iondna,iondna2}.
Accordingly, each Y-DNA arm is modeled as a chain of charged monomers consisting of a single ds-DNA junction monomer followed by twelve ds-DNA chain monomers and a single-stranded end group of four 
monomers. 
While the first thirteen monomers correspond to base pairs, the last four represent single nucleobases.
The connection between two Y-configurations is established by replacing four + four ss-DNA monomers with four ds-DNA monomers, see Fig.~\ref{fig:fig0}.
The numbers of constituents of each generation are given in Table~\ref{tab:tab1}.%
\begin{table}[h]
  \begin{center}
    \begin{tabular}{cccc} 
      Generation & $N_\mathrm{Y}$ &  $N_\mathrm{bp}$ & $N_\mathrm{mon}$ \\ \hline
      G1 & 1 & 39 & 51 \\
      G2 & 4 & 168 & 192 \\
      G3 & 10 & 426 & 474 \\
      G4 & 22 & 942 & 1038 \\
      G5 & 46 & 1974 & 2166 \\
      G6 & 94 & 3666 & 4422 \\ 
    \end{tabular}
  \end{center} 
  \caption{Characteristic numbers $N_\mathrm{Y}$ (i.e., number of Y-DNAs), $N_\mathrm{bp}$ (i.e., number of base pairs), and $N_\mathrm{mon}$ 
  (i.e., number of monomers) for DL-DNA of different generations.}
  \label{tab:tab1}
\end{table}

\begin{table}[h]
  \begin{center}
    \begin{tabular}{lcccccc} 
      Particle type  & Mass \(m\)[\si{\amu}] & Charge \(q\)[\(e\)] & Radius \(r_\alpha\)[\si{\angstrom}] \\ \hline
      regular monomer ($\rm M^-$) & 660 & -1 & 9 \\
      Y-junction monomer ($\rm M^-$) & 660 & -1 & 9 \\
      sticky end-linker ($\rm M^-$) & 330 & -1 & 9 \\
      counterions ($\rm C^+$) & 20 & +1 & 2 \\
      salt particles ($\rm S^\pm$) & 20 & \(\pm 1\) & 2 \\ 
    \end{tabular}
  \end{center}
  \caption{Properties of system's constituents. The radii at the last column refer to the model in eq.~(\ref{eq:lenjon}).}
  \label{tab:tab2}
\end{table}

The beads of the DNA-strands carry electric charges, mobile counterions are introduced in order to preserve the electroneutrality of the system.
Additionally, we also introduce different concentrations of monovalent salt ions, $\rm Na^+$ and $\rm Cl^-$, with the purpose of 
studying the influence of salt on the conformational characteristics of DL-DNA. 
The properties of each particle type can be seen in Table~\ref{tab:tab2}.
The steric interaction is described via truncated and vertically shifted Lennard-Jones potential, which is equivalent to the Weeks-Chandler-Andersen (WCA) potential, 
here with a possible horizontal shifting by $r_{\alpha\beta}$ as follows: 
  
\begin{equation}
  \label{eq:lenjon}
  V^{\alpha\beta}_\mathrm{steric}(r) =
  \begin{cases}
    \infty & \,\, \text{if } r < r_{\alpha\beta}\text{,} \\
    4 \epsilon \left[ \left( \frac{\sigma}{r - r_{\alpha \beta}} \right)^{12} - 
    \left( \frac{\sigma}{r - r_{\alpha \beta}} \right)^{6} + \frac{1}{4} \right] & \,\, \text{if } r_{\alpha\beta} \le r \le \sqrt[6]{2} \sigma + r_{\alpha\beta}\text{,} \\
    0 & \,\, \text{if } r > \sqrt[6]{2} \sigma + r_{\alpha\beta}\text{,} \\
  \end{cases}
\end{equation}
with the following parameters 
values: \(\sigma = \SI{4}{\angstrom} \),  \(\epsilon = 1\rm{kJ \cdot mol^{-1}}\)  
and  \(r_{\alpha\beta} = r_{\alpha} + r_{\beta} - \sigma\) (with \(\alpha, \beta = \mathrm{M}^-\), \(\mathrm{C}^+\), or \(\mathrm{S}^\pm\), referring to monomers, counterions, 
and counter/co-ions belonging to the salt particles, respectively), where \(r_{\mathrm{M}^-} = \SI{9}{\angstrom}\) and \(r_{\mathrm{C}^+} = r_{\mathrm{S}^\pm} = \SI{2}{\angstrom}\).  
This way, the excluded volume interaction between counterion particles reduces to the usual WCA interaction, diverging at zero separation, 
while the potential between monomers and ions diverges at a center-to-center distance of $\SI{7}{\angstrom}$ and the interaction between monomers diverges at the distance of $\SI{14}{\angstrom}$, accounting for the larger size of the monomers. Thus, the steric interaction acts in the range 
of  $r\leq \SI{18.5}{\angstrom}$, corresponding to the value of the effective DNA helix diameter, which is approximately equal $\SI{20}{\angstrom}$.

Consecutive monomers along ds-DNA- or ss-DNA-strands are connected with bonds described by a harmonic bonding potential:
\begin{equation}
  V_\mathrm{b}(r)=\frac{k_\mathrm{b}}{2}(r-l_\mathrm{b})^2,
\end{equation}
where the spring constant takes the value \(k_\mathrm{b} = \SI{210}{\kJ\per(\mol\cdot\angstrom^2)}\), giving a dispersion of about \SI{0.15}{\angstrom}, which is consistent with the values observed in structural studies of DNA~\cite{DNAb,DNAstruct}. 
The equilibrium bond length is \(l_\mathrm{b} = \SI{3.4}{\angstrom}\), 
(which renders the sping constant equivalent to $k_{\rm b}l_{\rm b}^2 \approx 10^3 k_{\rm B}T$)
corresponding to the typical distance between base pairs in DNA~\cite{DNAb,DNAm2}.
Comparing the bond length \(l_\mathrm{b}\) with the steric
monomer-monomer interaction offset \(r_{\mathrm{M^-}\mathrm{M^-}} = \SI{14}{\angstrom}\) reveals that neighbouring monomers in a straight configuration are located in the 
divergent regime of the WCA potential.
Therefore, we set the WCA steric interaction to only act between monomers which are \emph{not} within the same chain and additionally exclude the steric interaction between the first five monomers located at the Y-
DNA junctions.

The stiffness of ds-DNA is modeled via a harmonic bending-angle potential
$V_{\rm bend}(\phi)$, which acts on the angle $\phi$ between the bonds connecting any monomer (index \(j\)) and to the two neighbouring monomers (indices \(j + 1\) and \(j - 1\)):

\begin{equation}
  V_{\rm bend}(\phi)=\frac{k_{\phi}}{2}(\phi - \pi)^2, 
\label{vphi:eq}
\end{equation}
where the constant of bending energy, $k_\phi$, takes the values:
\begin{equation}
 k_\phi =
  \begin{cases}
    750 \ \rm kJ \cdot mol^{-1} & \quad \text{ for stiff chains,} \\
    150 \ \rm kJ \cdot mol^{-1} & \quad \text{ for sticky ends,} \\
    0 \  & \quad \text{ for the fully flexible Y-junction.} \\
  \end{cases}
\end{equation}
The bending energy constant value \(k_\phi = \SI{750}{\kJ \cdot \mol^{-1}}\) is chosen to reproduce the typical persistence length \(L = 500 - 1000 \si{\angstrom}\) of ds-DNA at low ionic strength~\cite{DNAl}.
While the bonds between the three central junction monomers are the same as between all other monomers, the Y-arms are fully flexible, i.e., the bending energy constant is chosen to be zero.
Since the persistence length of unpaired ss-DNAs is lower than for ds-DNAs~\cite{DNAss,pnas2012}, the degree of flexibility of the ssDNA end group monomer is set to \(k_\phi = \SI{150}{\kJ \cdot \mol^{-1}}\).

Since each monomer bears an elementary charge  \(q_\mathrm{M^-} = -e <0\), a corresponding number of counterions 
with charge $e > 0$ is added to ensure overall electroneutrality of the system. 
Any two charged species \(\alpha\) and \(\beta\) interact additionally via the Coulomb interaction
\begin{equation}
  \frac{V_{\mathrm{C}}(r)}{k_\mathrm{B}T} = \lambda_\mathrm{B}\frac{q_{\alpha}q_{\beta}}{r}\text{ ,}
\end{equation}
where \(r\) denotes interparticle separation, $q_{\alpha}, q_{\beta} \in \{e,-e\}$,
and the Bjerrum length \(\lambda_\mathrm{B} =\frac{e^2}{\epsilon_\mathrm{r} \epsilon_0 k_{\rm B} T}\) is set to \(\lambda_{\rm B} = \SI{7}{\angstrom}\). Water is treated as uniformly dielectric with dielectric 
constant \(\epsilon_\mathrm{r} = 80\).
\begin{figure}[h]
  \centering{\includegraphics[width=0.5\textwidth]{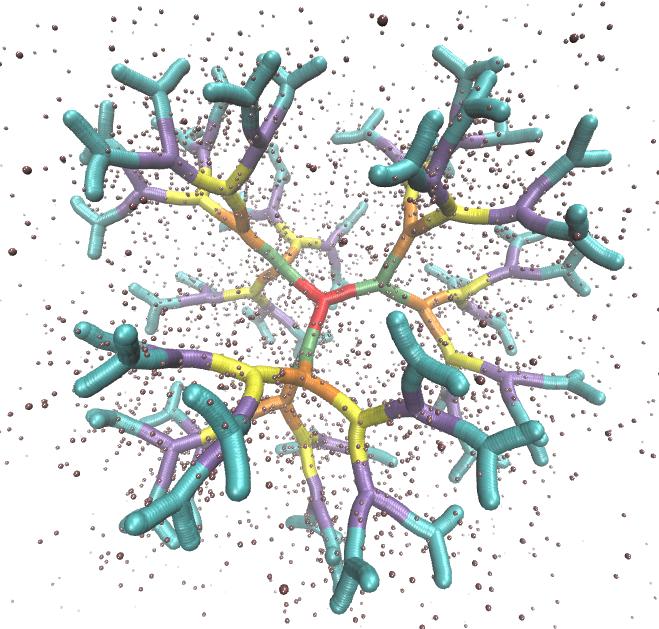}}
  \caption{Simulation snapshot of a G6 dendrimer. Each color corresponds to a different dendrimer generation. The figure  shows only a small fraction of simulation box. Small spheres represent counterions.}
  \label{fig:fig1}
\end{figure}

The above described model is used in Molecular Dynamics (MD) simulations, details of the latter are given in the {\bf Methods} section.
A representative simulation snapshot for a G6 dendrimer is shown in~Fig.~\ref{fig:fig1},
where monomers pertaining to different subgenerations are presented by different colors. 
It can be seen that the connections between successive branching points are rather rigid segments with a strong overlap between neighboring monomers, 
while the sticky ends (i.e., the segments pertaining to ssDNA belonging to the outermost subgeneration) show a less stiff behavior. 
Counterions are found to be absorbed into the interior of the dendrimer to a high degree.

\section{Results and discussion}

{\bf Comparison between experiment and simulation.} -- The overall size of the dendrimer can be characterized by its radius of gyration $R_\mathrm{g}$ or its hydrodynamic radius $R_\mathrm{H}$.  
In principle, these quantities can be determined in simulations;
they are also experimentally accessible by means of different scattering techniques, e.g., SANS, SAXS, or dynamic light scattering.
However, in this contribution, we only calculate the radius of gyration, $R_\mathrm{g}$, from simulated systems via the following expression:
\begin{equation}
  \label{eq:eq6}
  R_\mathrm{g}^2 = \frac{1}{N} \left< \sum_{i=1}^N \left(\bm{r}_i - \bm{r}_\mathrm{com} \right)^2 \right> \,\text{,}
\end{equation}
where \(N\) is the total number of monomers constituting the DL-DNA, \(\bm{r}_i\) denotes the positions of the individual monomers, and \(\bm{r}_\mathrm{com}\) stands for the center of mass of the molecule. 
This quantity is readily accessible in MD simulations and enables us to assign a typical size to the molecule described by employed model.  
The hydrodynamic radius \(R_\mathrm{H}\) was experimentally determined using dynamic light scattering and measuring the diffusion coefficient in diluted dendrimer solutions. 
Though the two radii are different by definition, one measuring spatial extent and the other hydrodynamic drag, they differ in their values only by a small amount
so that a comparison of $R_\mathrm{g}$ from simulation with \(R_\mathrm{H}\) from experiment is a good way to validate the model.
Moreover, static light scattering has been also employed to determine $R_\mathrm{g}$ for dendrimers of higher generation numbers.

\begin{figure}[h]
  \begin{center}
    \includegraphics[width=0.5\textwidth, angle = -90]{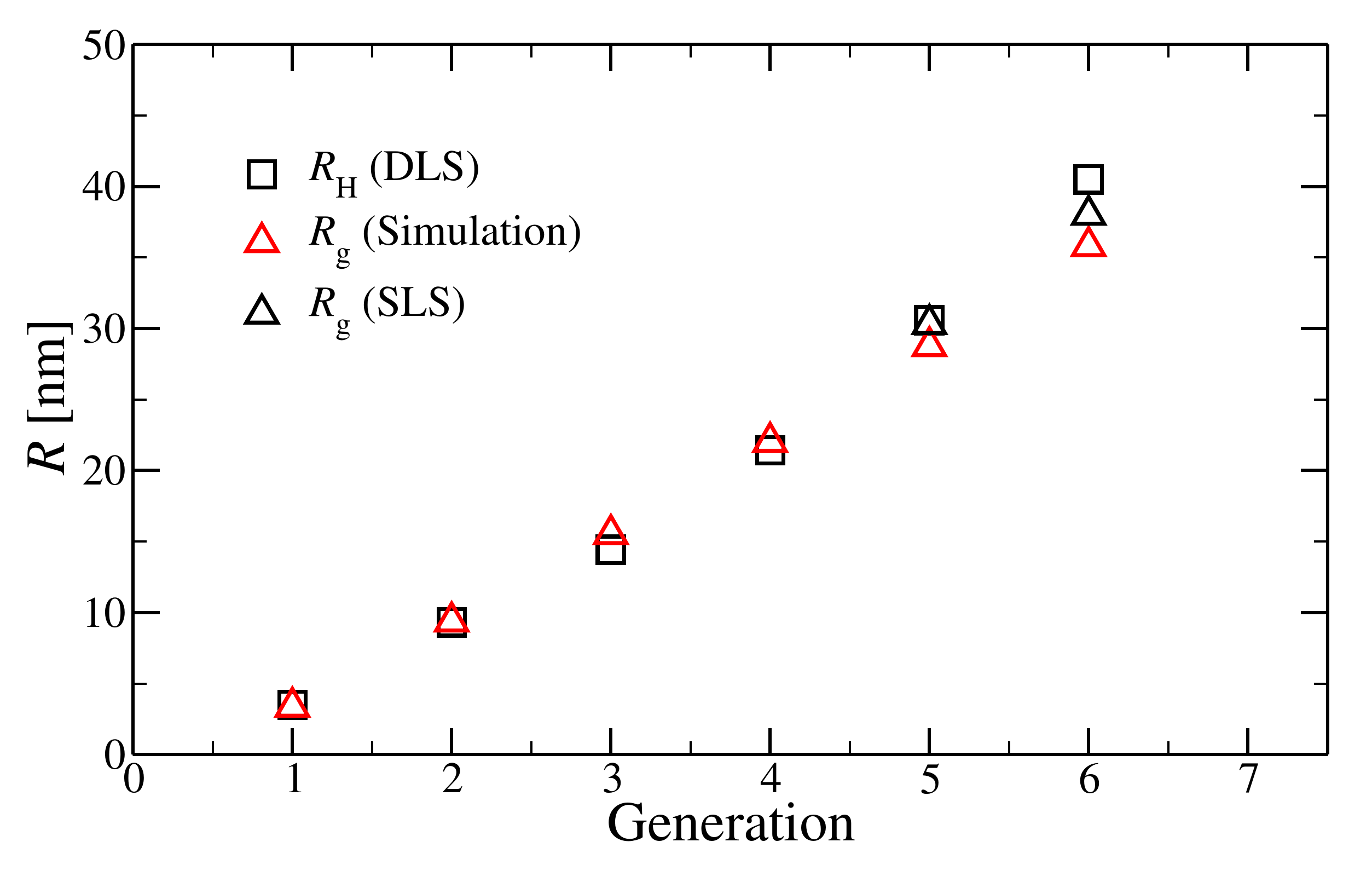}
    \caption{
      (color online) Hydrodynamic radius \(R_\mathrm{H}\) (as extracted from experiment) and radius of gyration \(R_\mathrm{g}\) (as predicted in simulation) 
      as functions of the generation index of DL-DNAs (as labeled). For DL-DNA of G5 and G6 experimental value of $R_{\rm g}$ is also provided.} 
          \label{fig:plt_rg_all}
  \end{center}
\end{figure}

A comparison of the results originating from experiment, \(R_\mathrm{H}\), and simulation, \(R_\mathrm{g}\), is presented in Fig.~\ref{fig:plt_rg_all}.
For G1 to G5 the results of \(R_\mathrm{g}\) and \(R_\mathrm{H}\) show excellent agreement, indicating the appropriateness of the underlying model. 
The two sets of data show small discrepancies for DL-DNAs of higher generations, i.e., for G6 DL-DNAs;
therefore, we show here also experimental values of $R_{\rm g}$ for G5 and G6 DL-DNAs. 
As experiments show, for G5 DL-DNAs radius of gyration coincides with hydrodynamic radius $R_{\rm g}=R_{\rm H}$, which on the other hand, fits nicely with the obtained simulation value.
This matching between results of static and dynamic light scattering  for G5 justifies our choice of comparing two different quantities that characterize the size of a dendrimers of lower generation numbers.
We observe that both the experimental and simulation data exhibit a concave shape as function of generation index reflecting the non-linear growth of the dendrimer with increasing generation number.
This feature can be explained by the observed increase of the molecules' sphericity with the growing monomer density at the periphery of the DL-DNA. 
As the sphericity of the dendrimer increases, as it is the case of G6 DL-DNAs, the ratio between the experimentally measured radius of 
gyration and the hydrodynamic radius deviates from 1  and goes toward smaller values, i.e. $R_{\rm g}/R_{\rm H} = 0.94$. 
For the sake of comparison, the theoretical value of this ratio is 0.778 for a homogeneous hard sphere~\cite{Richter} and 1.0 for hollow spheres
with a infinitely thin shell~\cite{vesicle}.
Therefore, the significant discrepancy observed between the experimental $R_{\rm H}$ and $R_{\rm g}$ obtained from simulation for 
G6 DL-DNAs and probably also for higher generations becomes reasonable and one has to employ static light scattering in order to obtain better agreement with the results of performed type of simulations. 

{\bf Conformational analysis.} -- A more detailed analysis of the form factor \(F_\mathrm{mm}(q)\) of the dendrimer provides a deeper 
insight into the structural properties of the dendrimer. The form factor is defined as
\begin{equation}
F_\mathrm{mm}(q) = 1 + \frac{1}{N} \left\langle \sum_{i\neq j}^N \exp{[-i \left(\bm{q} \cdot \bm{r}_{ij}\right)]} \right\rangle \,\text{,}
\end{equation}
where the summation runs over all inter-monomer distances \(\bm{r}_{ij}\) and the brackets $\langle \ldots \rangle$
stand for an average over all conformations, which restores rotational symmetry; here, \(\bm{q}\) denotes the scattering wavevector,
allowing to look at different scales within the molecule. At coarse length scales, \(q R_\mathrm{g} \lesssim 1\), 
the above expression reduces to the Guinier law~\cite{guinier}:
\begin{equation}
  \label{eq:eq7}
  F_\mathrm{mm}(q) \simeq N \exp{\left[\frac{-\left(q R_\mathrm{g}\right)^2}{3}\right]} \,\text{,} 
\end{equation}
which represents a useful relation between the form factor and the radius of gyration within the regime \(q R_\mathrm{g} \lesssim 1\).
Via this expression, the radius of gyration can be extracted from experimental form factor data in the small wave-vector limit.

\begin{figure}
\centering{\hspace{-1cm}\begin{subfigure}{.33\textwidth}\includegraphics[width=0.687\textwidth, angle=270]{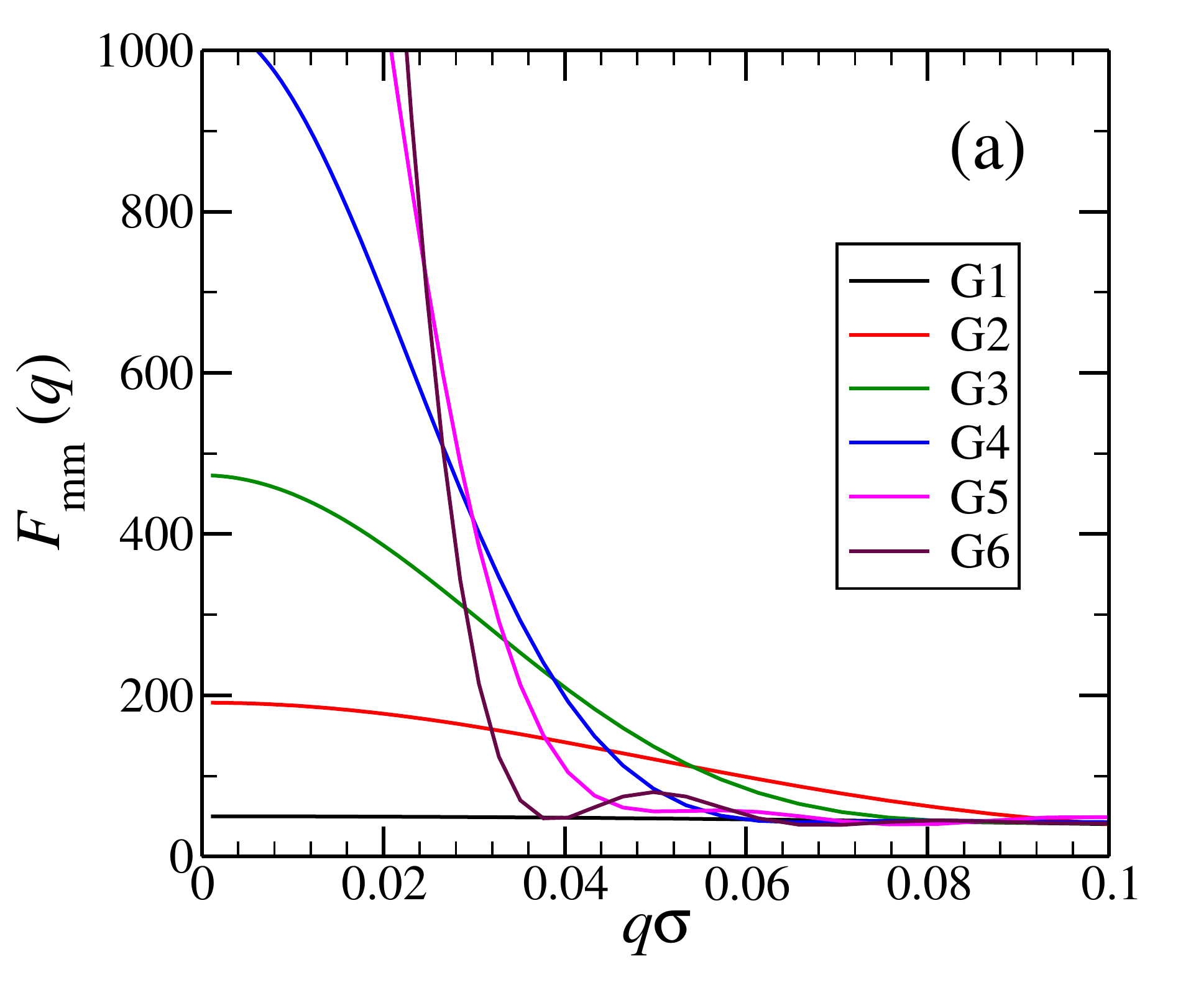}\end{subfigure}\hspace{-1.2cm}\begin{subfigure}{.33\textwidth}
\includegraphics[width=0.687\textwidth, angle=270]{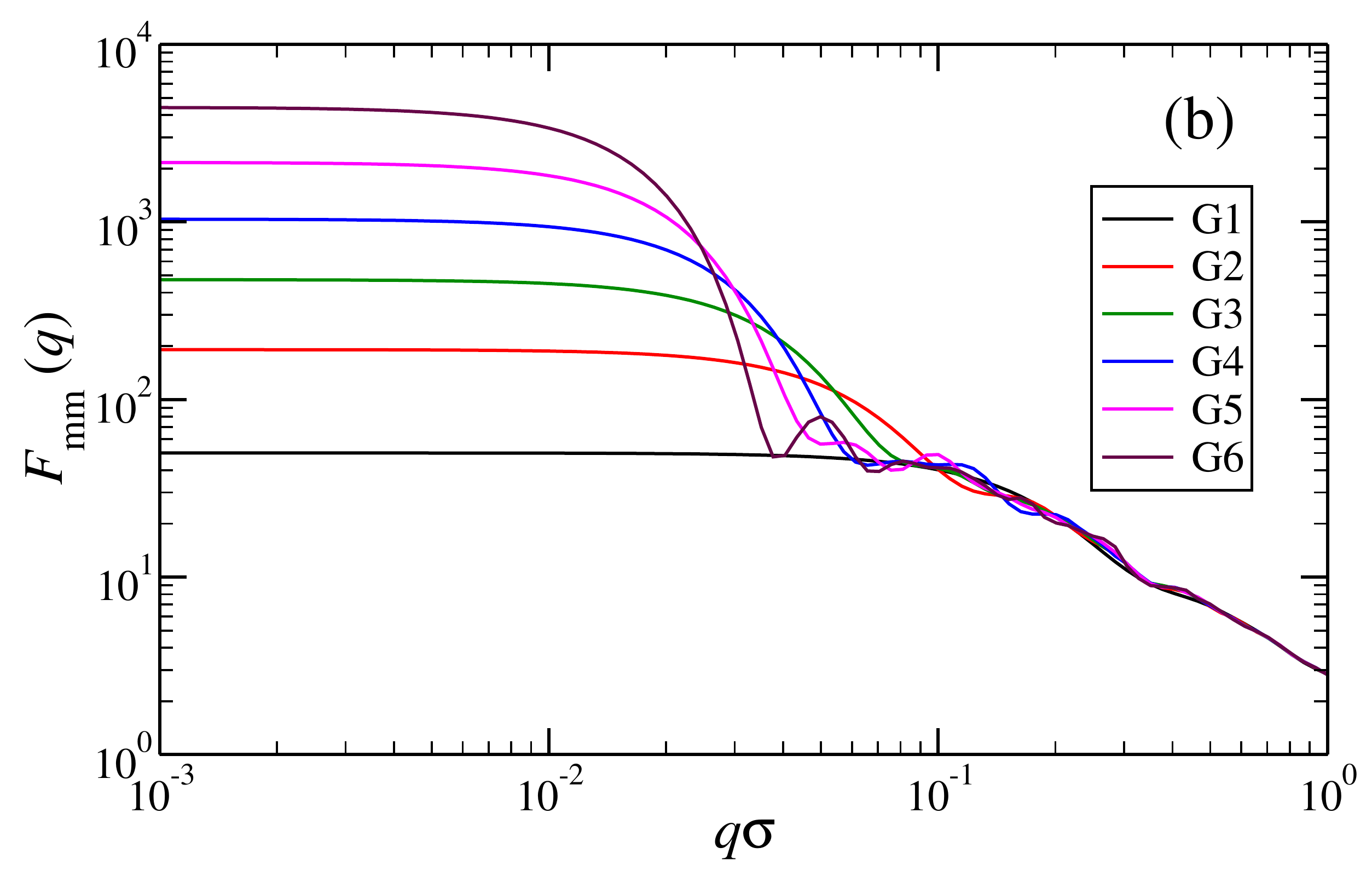}\end{subfigure}\hspace{0.71cm}\begin{subfigure}{.33\textwidth}\includegraphics[width=0.687\textwidth, angle=270]{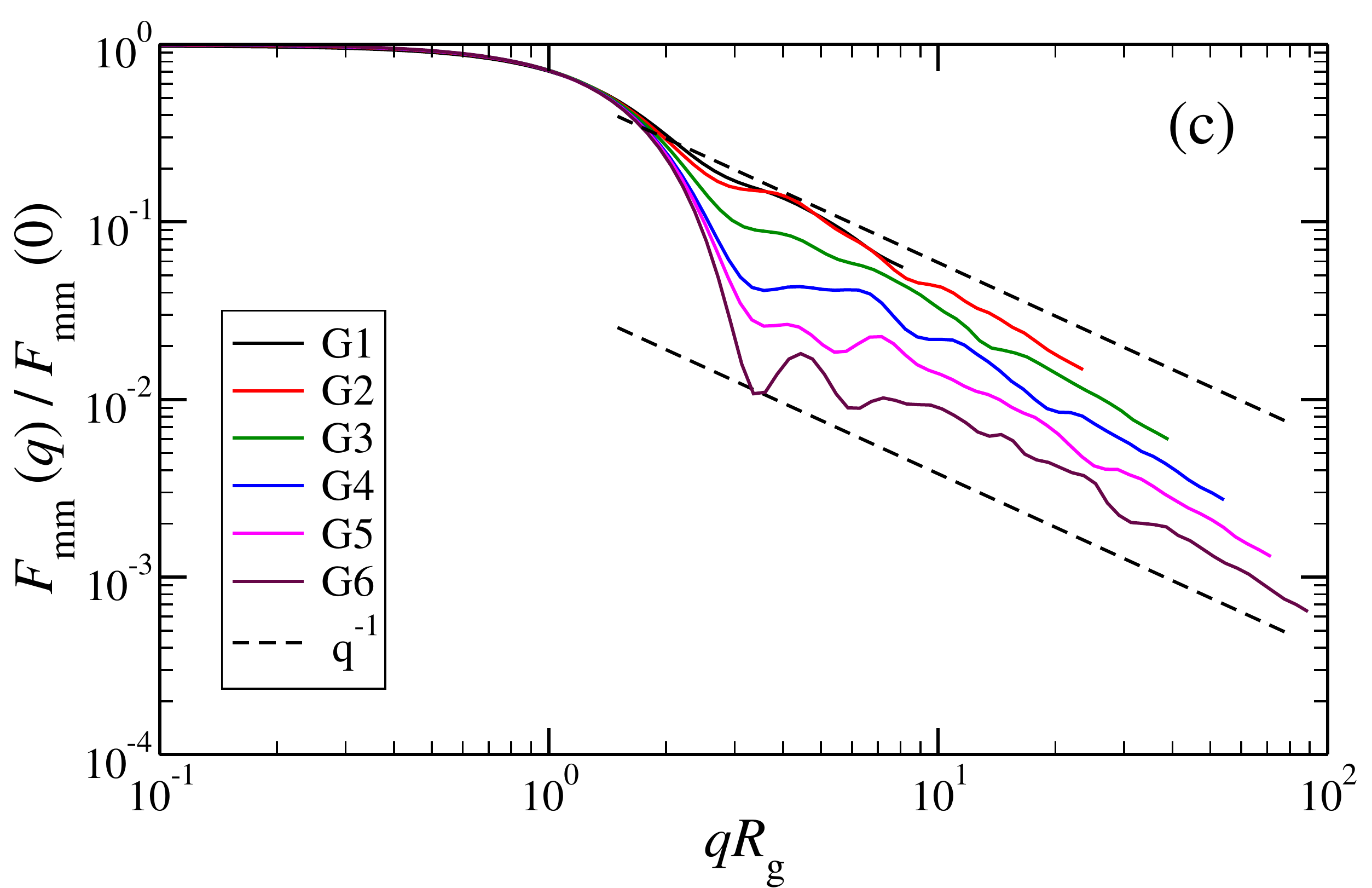}\end{subfigure}}
    \caption{The form factor \(F_{\mathrm{mm}}(q)\) of the DL-DNAs of generation G1 to G6 (as labeled) obtained from simulations, given as a function of dimensionless wave-vector. 
      Data are shown: (a) on a linear scale; (b) on a double-logarithmic scale; and (c) on a double-logarithmic scale with \(\lim_{q \to 0}F_{\mathrm{mm}}(q)\) rescaled to \(1\).
      The dashed lines in panel (c), i.e., \(F_\mathrm{mm}(q) \sim q^{-1}\), corresponds to the typical scaling law for the scattering from rigid rods~\cite{rods} for large \(q\)-values.}
    \label{fig:fig5}
\end{figure}

An overview of the results for \(F_\mathrm{mm}(q)\) from simulation is given in Fig.~\ref{fig:fig5}, where this function is shown on different scales.
As eq.~\eqref{eq:eq7} implies, the form factor becomes equal to the total number of monomers of the molecule in the limit of small wave vectors, i.e., for \(q \to 0\).
Further, we observe oscillations in \(F_\mathrm{mm}(q)\) for \(q\sigma \sim 10^{-1}\);
the first local minimum becomes more pronounced with increasing generation index, signifying that these larger 
molecules possess a more spherical shape and that the sharpness of the molecules' boundary at the outermost shell increases.
The rigidity of the ds-DNA strands within the molecule reflects in the large wave-vector behaviour of \(F_\mathrm{mm}(q)\),
namely, the form factor satisfies the law \(F_{\rm mm}(q) \sim q^{-1}\) in the limit of large wave vectors~\cite{rods}.
This is the typical scaling law for scattering from rigid rods, which is in contrast to flexible dendrimers, which usually scale with \(\sim q^{-4}\) 
in the same range of $q$, according to Porod's law~\cite{porod}.
This outcome can be better understood in the context of  scattering from fractal aggregates, it can be shown~\cite{scat-frac} that for arbitrary systems of scatterers 
the scattering intensity $F_{\rm mm}(q)$ scales with the wave 
vector as
\begin{equation}
 F_{\rm mm}(q) \sim (qR)^{-2 D_\mathrm{m} + D_\mathrm{s}} \quad \text{for } R^{-1} < q < a^{-1} \text{,}
\end{equation}
where \(D_\mathrm{m}\) and \(D_\mathrm{s}\) are the mass and surface fractal dimensions, respectively. The size of a single monomer is \(a\), while the size of the whole system is denoted as \(R\).
In the case of solid spheres the mass dimension \(D_\mathrm{m}\) is equal to the system's spatial dimension \(d\) (\(D_\mathrm{m} = d = 3 \)), while the surface dimension is \(D_\mathrm{s} = d - 1 = 2\), 
which results in the well-known Porod's law, \(F_{\rm mm}(q) \sim q^{-4}\). 
On the other hand, a rigid dendrimer can be characterized as a single fractal aggregate. 
Therefore, its fractal dimension, \(D\), is equal to the fractal mass and surface dimensions, \(D = D_\mathrm{m} = D_\mathrm{s} < d\), obeying the scaling law 

\begin{figure}[h]
  \begin{center}
        \includegraphics[width=0.4\textwidth, angle=270]{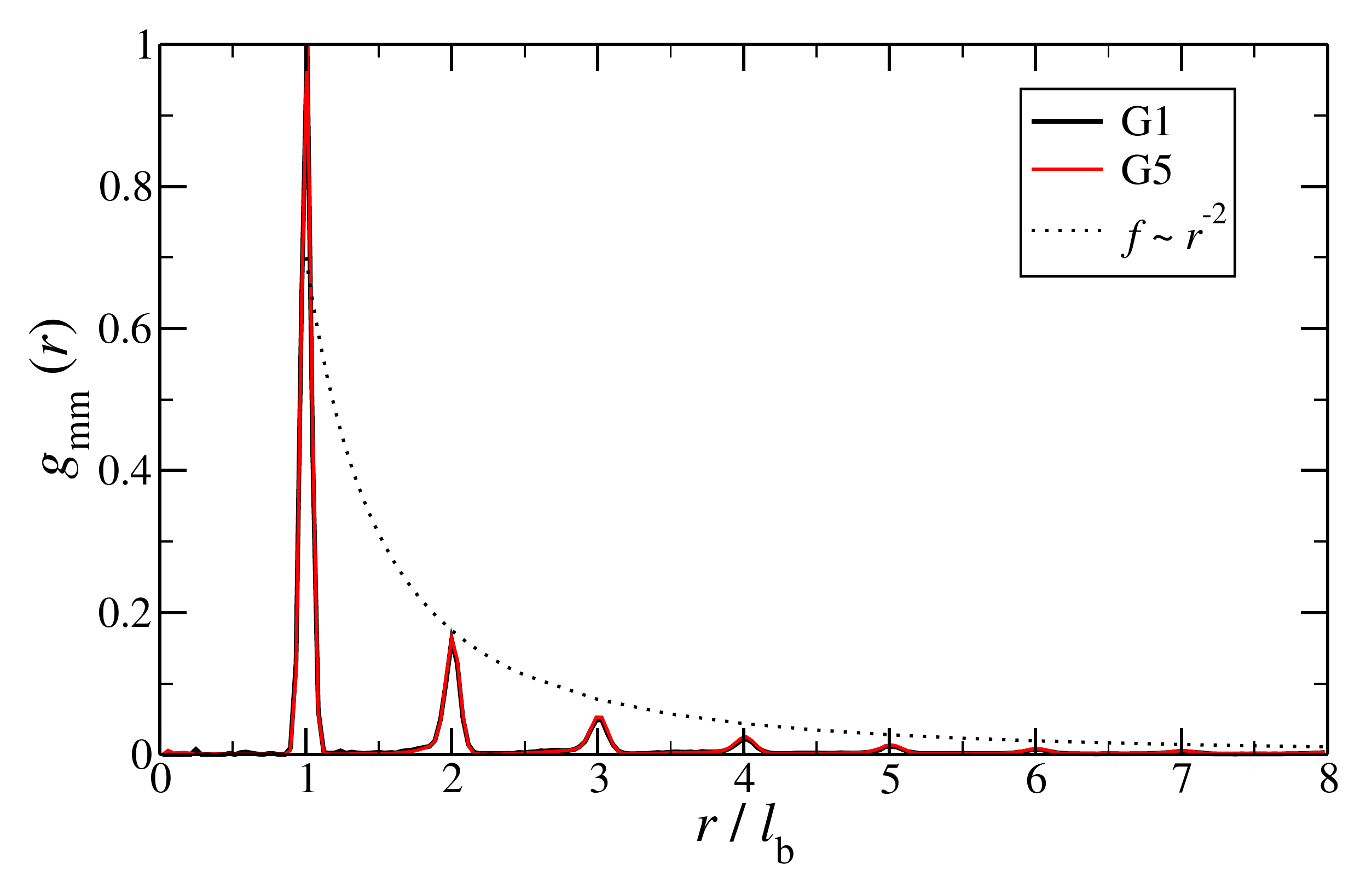}
    \caption{
      (color online) Monomer-monomer radial distribution function \(g_{mm}(r)\) for DL-DNAs of generations G1 and G5 plotted as functions of distance \(r\), given in units of the equilibrium bond length~\(l_\mathrm{b}\).
    }
    \label{fig:rdf}
  \end{center}
\end{figure}

\begin{equation}
F_{\rm mm}(q) \sim (qR_\mathrm{g})^{-D} \quad \text{for } qR_\mathrm{g} > 1,
\end{equation}
with \(D = 1\).
By analyzing the form factor in this way, we have obtained a beautiful reflection of the distinctive behaviours of flexible and rigid dendrimers. 

Additional insight into the conformational features of DL-DNAs can be acquired by analyzing the monomer-monomer pair correlation function.
In Fig.~\ref{fig:rdf} the radial distribution function \(g_\mathrm{mm}(r)\) for DL-DNAs of generations \(\mathrm{G}1\) and \(\mathrm{G}5\) is plotted as a function of the distance \(r\), 
given in units of the equilibrium bond length \(l_\mathrm{b}\).
The well-defined maxima which occur at equidistant positions indicate that the bonds between the monomers are rather stiff.
The first and largest peak represents the nearest neighbour separation along the Y-DNA arms.
The height of the maxima scales with \(r^{-2}\), which is the rate at which the volume of the spherical shells increases. 
The two curves of \(g_\mathrm{mm}(r)\) are identical for \(\mathrm{G}1\) and \(\mathrm{G}5\) in this regime.

\begin{figure}[h]
  \begin{center}
\centering{\begin{subfigure}{.5\textwidth}\includegraphics[width=0.6\textwidth, angle=270]{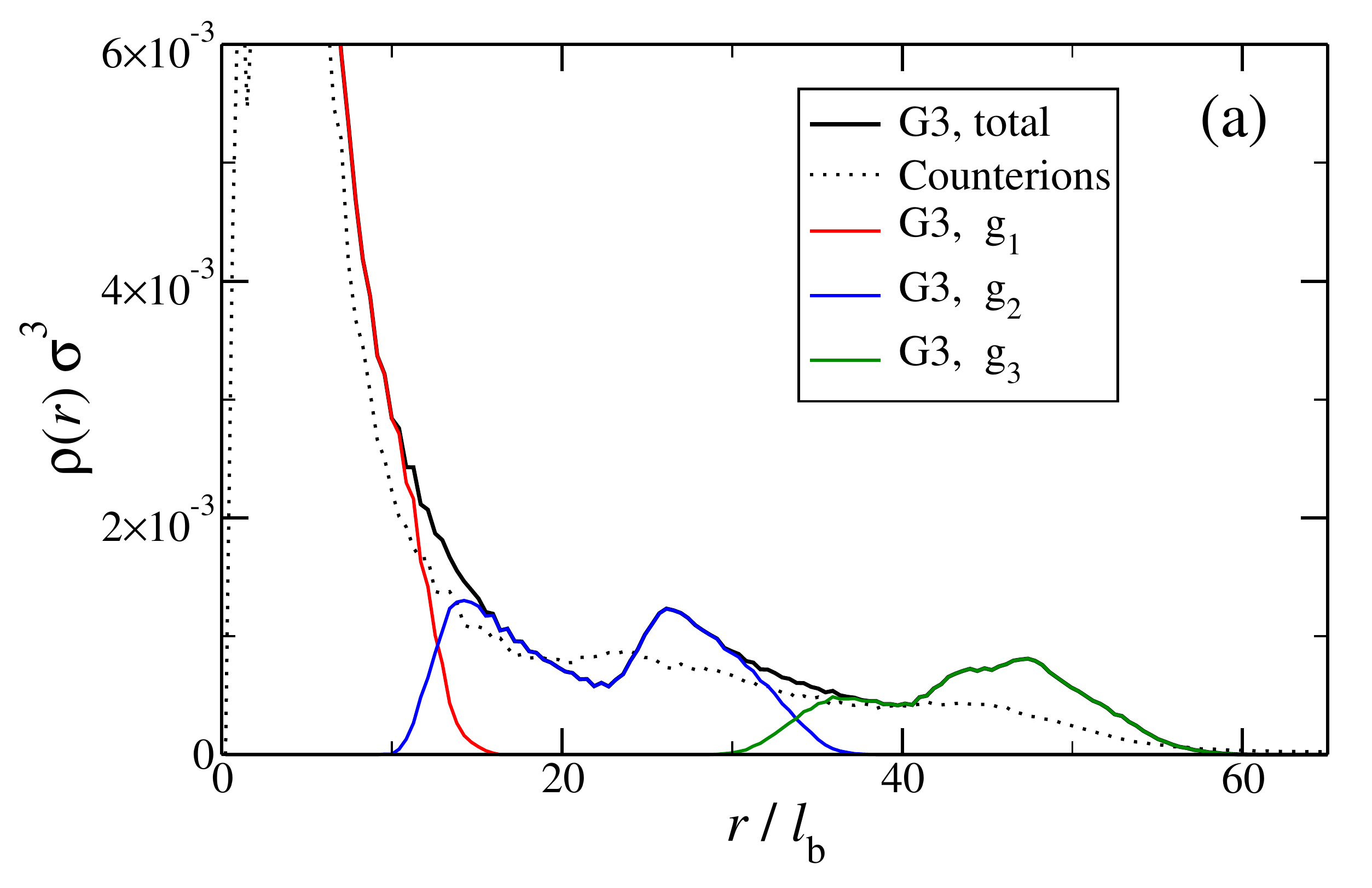}\end{subfigure}\begin{subfigure}{.5\textwidth}\includegraphics[width=0.6\textwidth, angle=270]{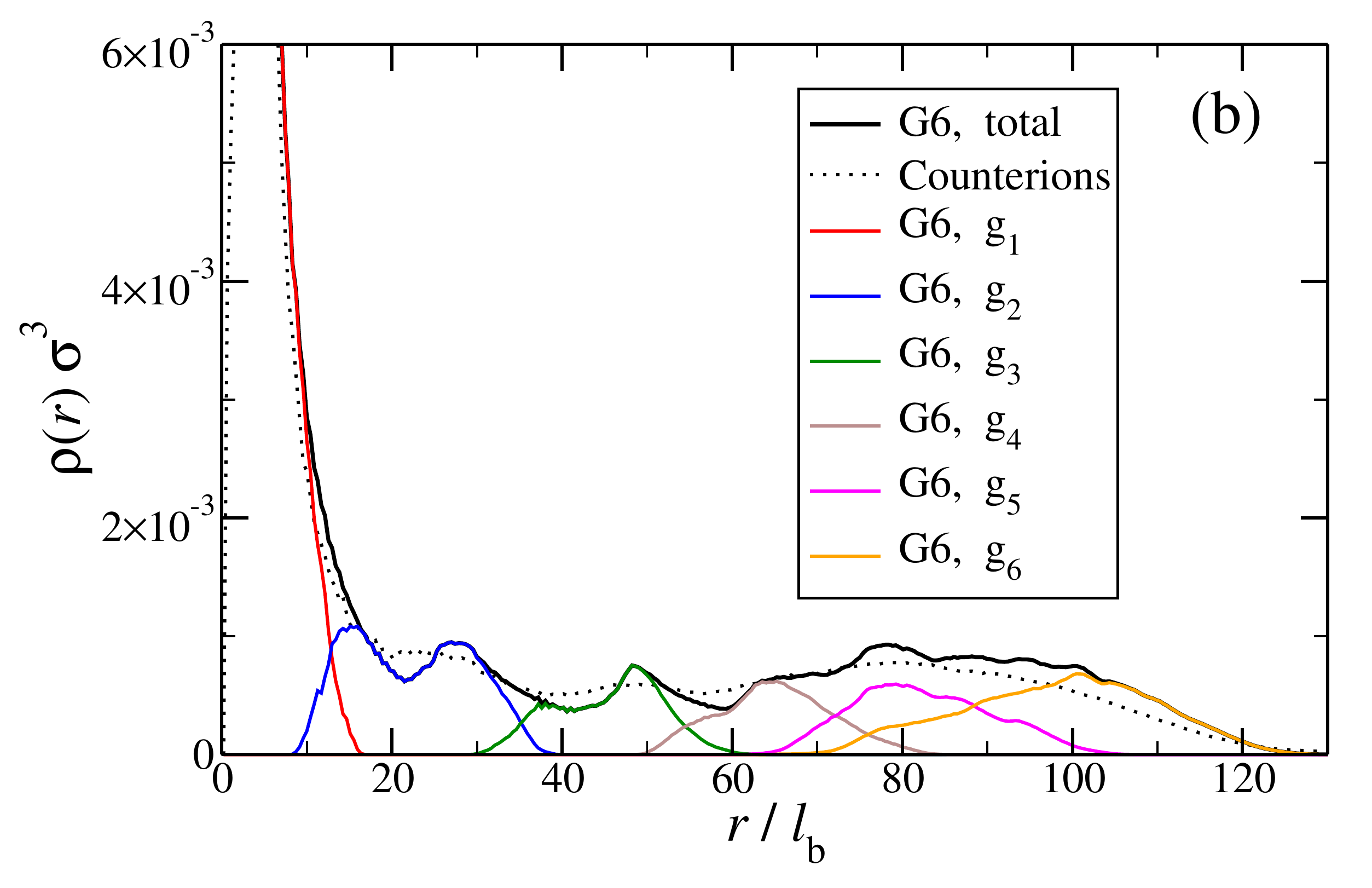}\end{subfigure}}
    \caption{
      (color online) Monomer and counterion density profiles \(\rho(r)\) as functions of \(r\), given in units of the equilibrium bond length \(l_\mathrm{b}\);
      \(\rho(r)\) is shown for different entities of the system (as labeled and see text).
      Data are shown (a) for  a G3 DL-DNA and (b) for a G6 DL-DNA.}
    \label{fig:fig3}
  \end{center}
\end{figure}

Another quantity that provides detailed information about the complex internal structure of DL-DNAs is the density profile of the constituents of the macromolecule with respect to the center of mass of the dendrimer:
\begin{equation}
  \rho(r) = \left\langle\sum_{i=1}^{N_{m}}\delta(\bm{r} - \bm{r}_i + \bm{r}_\mathrm{com}) \right\rangle,
\end{equation}
where the summation runs over all particles of a particular type, such as the monomers or the counterions; the vectors \({\bm r}_{i}\) denote the corresponding positions of the particles. 
In Fig.~\ref{fig:fig3} the density profiles for specific components of DL-DNA molecules are shown, focusing on a G3 (Fig.~\ref{fig:fig3}(a)) and a G6 (Fig.~\ref{fig:fig3}(b)) dendrimers, respectively. 
The different entities considered are (i) all monomers (without distinction; ``total''), (ii) the monomers pertaining to a specific subgeneration, \(\mathrm{g}_i\), and (iii) the counterions.
The monomers are regularly distributed in concentric-like structures such that only minor overlap between subsequent subgenerations exists. 
This kind of behavior is typical for charged macromolecules with rigid bonds~\cite{RB08,RB10}: the strong Coulomb repulsion, combined with rigidity and the
dendritic character
prevent backfolding of the outer monomers towards the interior of the molecule, a feature that is in contrast to the standard dense-core model of dendrimers with flexible bonds~\cite{dc1,dc2}. 
This rigidity is in addition reinforced by the Coulomb repulsion between like-charged monomers, resulting in a complete suppression of backfolding.  
Because the Y-DNAs of the individual subgenerations exhibit a transition at the junction, where one inward-facing arm splits into two outward facing arms, the corresponding 
density profiles feature a double-peak.
This double-peak feature is not as pronounced in subgenerations with higher index \(\mathrm{g}_i\), as the spatial distribution of these higher subgenerations is less coherent and more flattened out.
The counterion density distribution closely follows the monomer density due to the system's propensity towards local charge neutrality and the
spatial structure of the counterions is less pronounced due to an entropic `smearing out' of the profiles.
Overall, we obtain, especially for higher generations, almost
flat-density molecules, whose monomer-
and counterion-profiles are tunable by varying the generation index, \(\mathrm{G}N\).
These `uniform-density' nanostructures with a huge amount of free space in their interior
(see the vertical scale in Fig.~\ref{fig:fig3}) are suitable for an analytical description using the Poisson-Boltzmann theory,~\cite{klos, levin} since the 
constant ion density inside the molecule simplyfies analytical calculations. \newline

\begin{figure}[h]
  \begin{center}
\centering{\begin{subfigure}{.5\textwidth}\includegraphics[width=0.6\textwidth, angle=270]{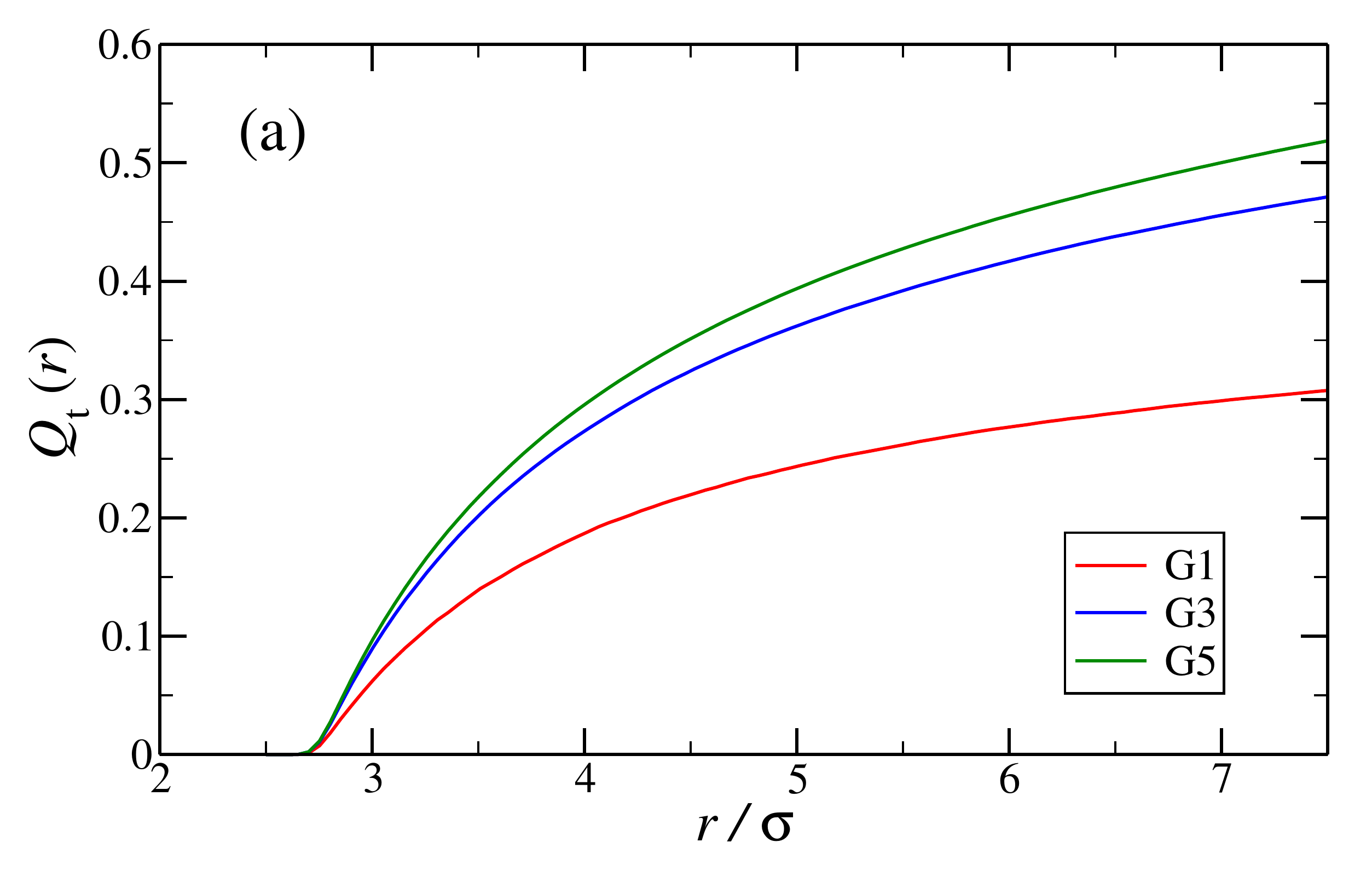}\end{subfigure}\begin{subfigure}{.5\textwidth}\includegraphics[width=0.6\textwidth, angle=270]{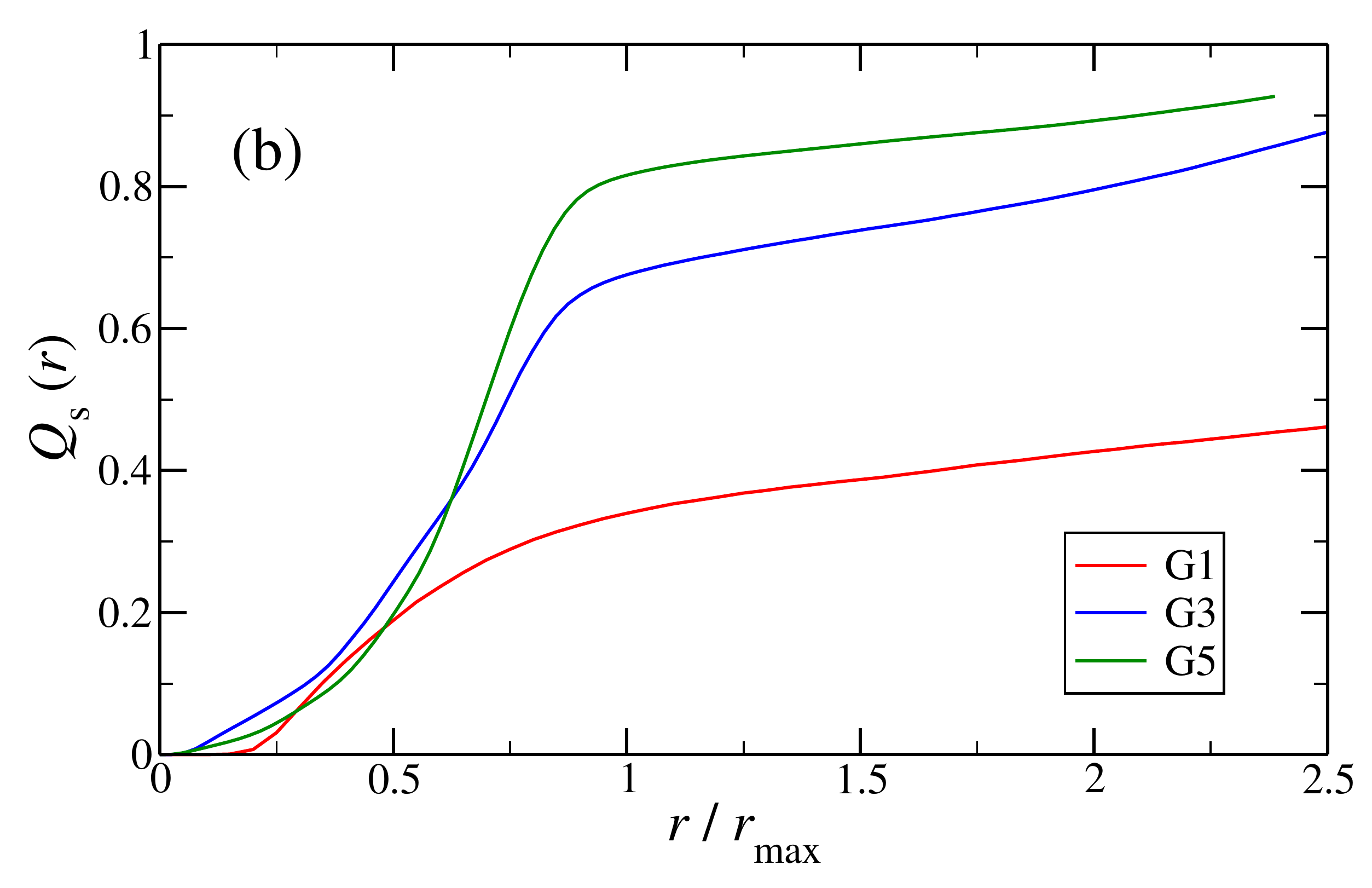}\end{subfigure}}
  \end{center}
  \caption{
    (color online) (a) Percentage of total counterions, \(Q_\mathrm{t}(r)\), captured in tubes of radius \(r\) surrounding each arm of the Y-DNA elements as a function of \(r\) (in units of \(\sigma\)).
    (b) Percentage of total counterions, \(Q_\mathrm{s}(r)\), captured in spheres of radius \(r\) centered at the center of mass of the DL-DNA \(\bf{r}_\mathrm{com}\) as 
    a function of \(r / r_\mathrm{max}\), where \(r_\mathrm{max}\) denotes the maximum distance between a DNA-monomer and \(\bf{r}_\mathrm{com}\).
    The graphs are shown for dendrimers G1, G3, and G5 (as labeled).}
  \label{fig:fig34}
\end{figure}

In the following, a more detailed analysis of the counterion condensation is presented.
In our investigation we have encircled each arm of the individual Y-DNA elements by a tube of radius \(r\); 
we have then counted the percentage of ions \(Q_\mathrm{t}(r)\), captured in those cylinders.
The dependence of \(Q_\mathrm{t}(r)\) on the tube radius \(r\) is depicted in Fig.~\ref{fig:fig34}(a).
Even though the considered system is electro-neutral, a difference in the value of \(Q_\mathrm{t}(r)\) of approximately \(20\%\) 
between G1 and G5 dendrimers can be observed for tube radii larger than \(3 r_{\mathrm{M}^-\mathrm{C}^+}\).
This observation is a direct consequence of an increase in the available volume provided by the larger dendrimers.
The same effect can be seen in Fig.~\ref{fig:fig34}(b), where the total amount of counterions absorbed by the dendrimers is shown, expressed via function \(Q_\mathrm{s}(r)\). When the radius of the sphere $r$ that encircles the dendrimer, exceeds the 
size of the dendrimer $r_{\rm max}$ only $40\%$ of counterions is absorbed by a G1 DL-DNA, while with the increase of dendrimer generation number that percentage grows and it approaches $90\%$ in the case 
of a G5 DL-DNA. It is also worth noticing that the transition of the counterion profile from the interior to the exterior becomes increasingly sharp as the 
dendrimer generation grows: accordingly, high-G DL-DNA's act as osmotic dendrimers, in full analogy with the osmotic polyelectrolyte stars~\cite{jusufi:prl:02,jusufi:jcp:02, DNAstar},
which capture the counterions in their interior. However, in contrast to these, DL-DNA's are very robust against salinity, maintaining their size essentially unaffected by addition
of large quantities of monovalent salt, as it will be shown in what follows.

\begin{figure}[h]
  \begin{center}
        \includegraphics[width=0.36\textwidth, angle = -90]{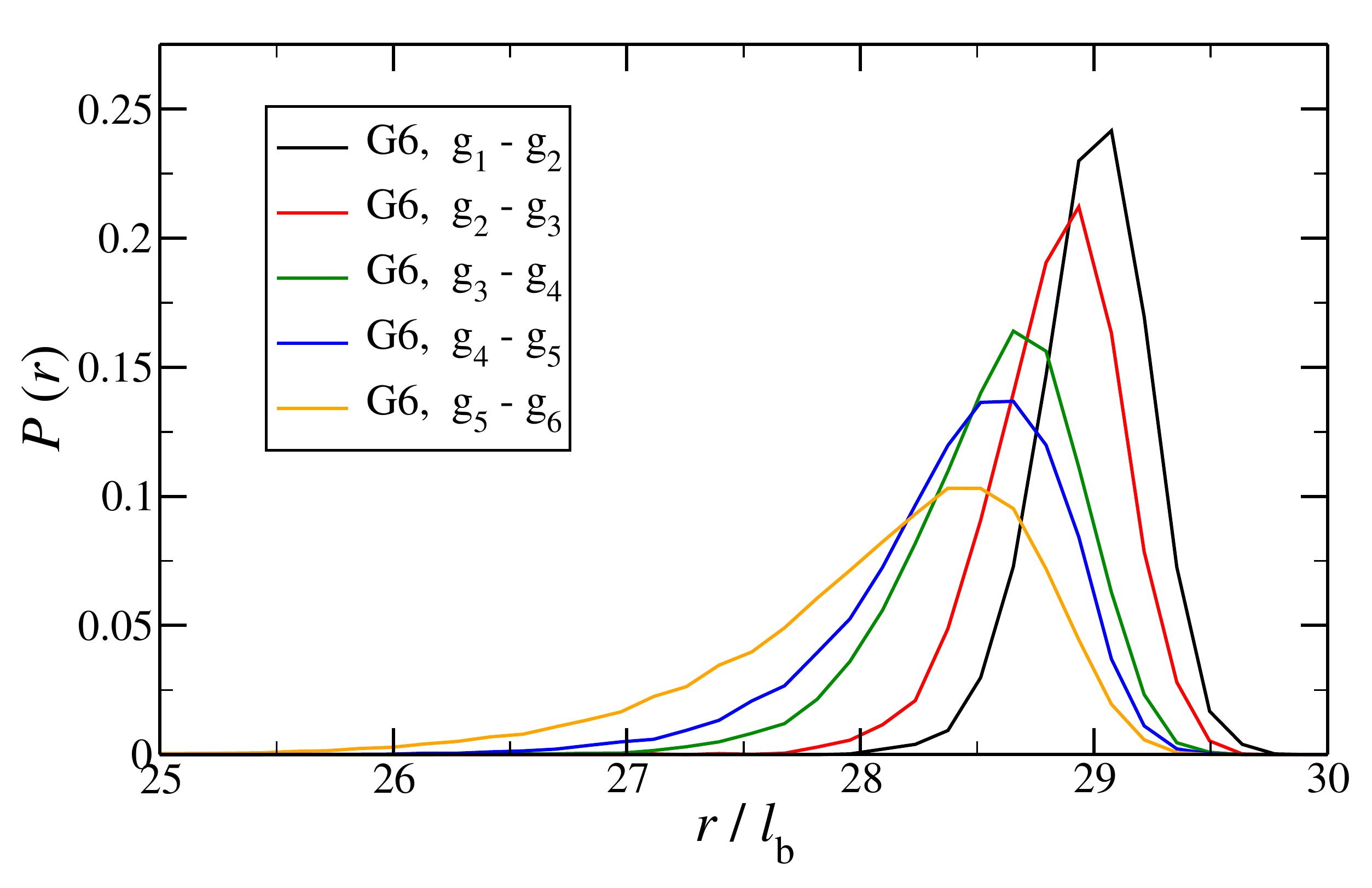}
    \caption{
      (color online) Probability for the nearest junction-to-junction separation of successive subgenerations $g_i$ and $g_{i+1}$ within a G6 DL-DNA given in units of the equillibrium bond length $l_{\rm b}$.
         }
    \label{fig:fig7}
  \end{center}
\end{figure}

In Fig.~\ref{fig:fig7}  the probability of the nearest junction-to-junction separation of successive subgenerations $g_i$ and $g_{i+1}$ within a G6 DL-DNA is shown as a function of distance given in units of bond length. 
Since each branch extending from junction to junction point consists of 30 monomers, the peak of this function is located between $28l_{\rm b}$ and $29l_{\rm b}$ for the innermost branches and its position decreases monotonically as  
one moves away toward the exterior of the molecule, (i.e., towards the outer subgroups).
The distributions that describe the outer branches are shifted for approximately half a bond length and these functions cover a wider range of distances.
The shrinkage of bond lengths belonging to the outer branches is a consequence of osmotic swelling which  tends, on one hand, to stretch central (inner) branches, while on the other hand, it allows 
a slightly higher 
flexibility of the branches belonging to higher subgenerations. To understand the physics behind this, we need to consider the osmotic pressure from the counterions trapped
in the interior of the molecule, which tries to swell the dendrimer by exercising an outward force at a ficticious spherical surface of radius $r_{\rm max}$ that surrounds the 
molecule, touching the free tips of the outermost Y-junctions (composed of 4-bases long ssDNA). This force is transmitted to the interior of the dendrimer but the number of Y-junction tips among
which it is partitioned is halved each time the generation index decreases by one. Accordingly, the innermost generations are pulled more strongly than the outermost ones;
thus they are thus more rigid, more straight-line looking.  
This effect is also observable in the simulation snapshot (see Fig.~\ref{fig:fig1}) where it
can be seen that the innermost branches are more rigid, having the shape of a straight line, while the branches belonging to the 
outermost subgenerations expose a more wiggly behavior. 
This interpretation is corroborated by the analysis of angular fluctuations in what follows.

\begin{figure}[t]
  \begin{center}
\centering{\begin{subfigure}{.5\textwidth}\includegraphics[width=0.6\textwidth, angle=270]{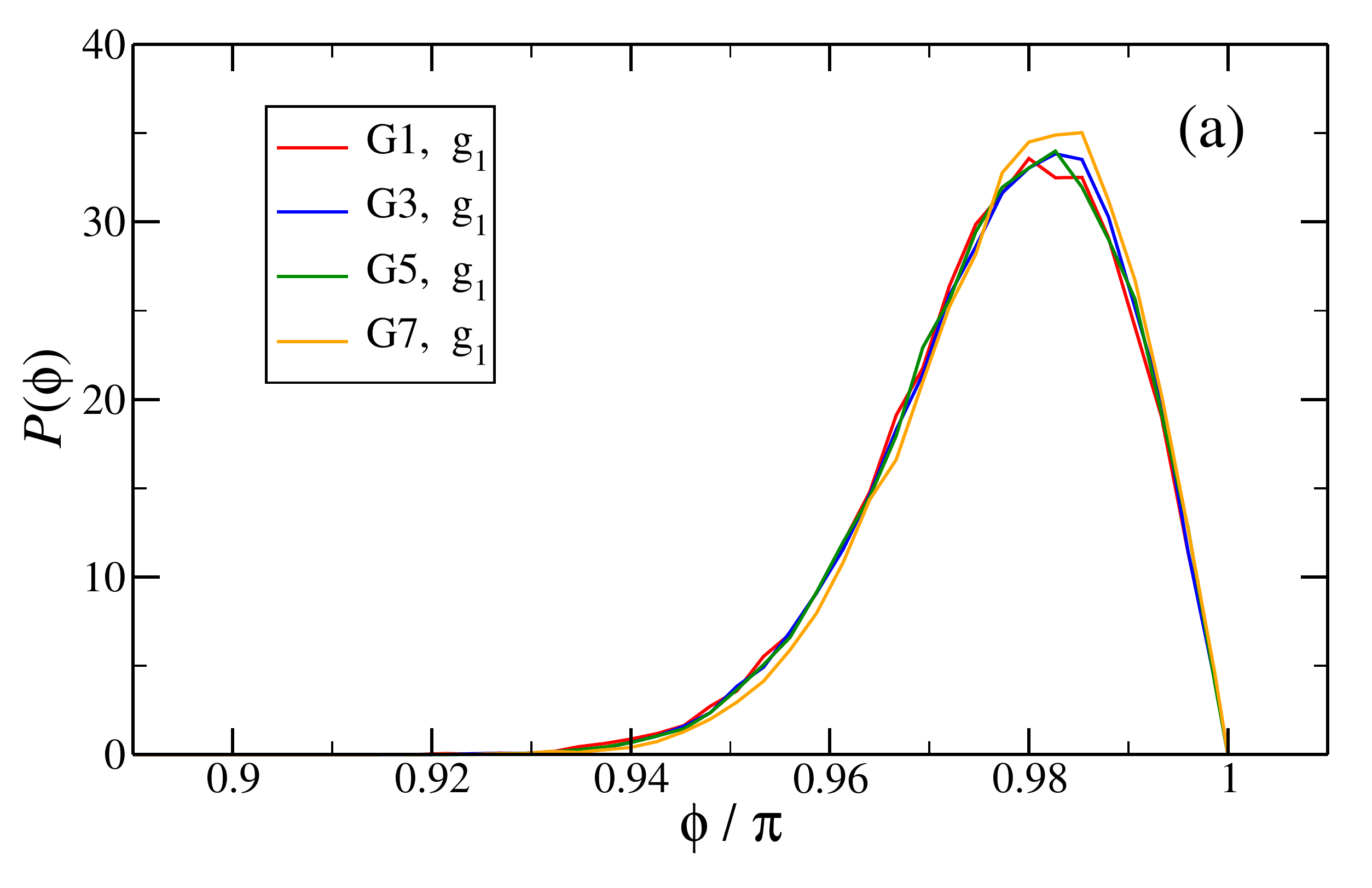}\end{subfigure}\begin{subfigure}{.5\textwidth}\includegraphics[width=0.6\textwidth, angle=270]{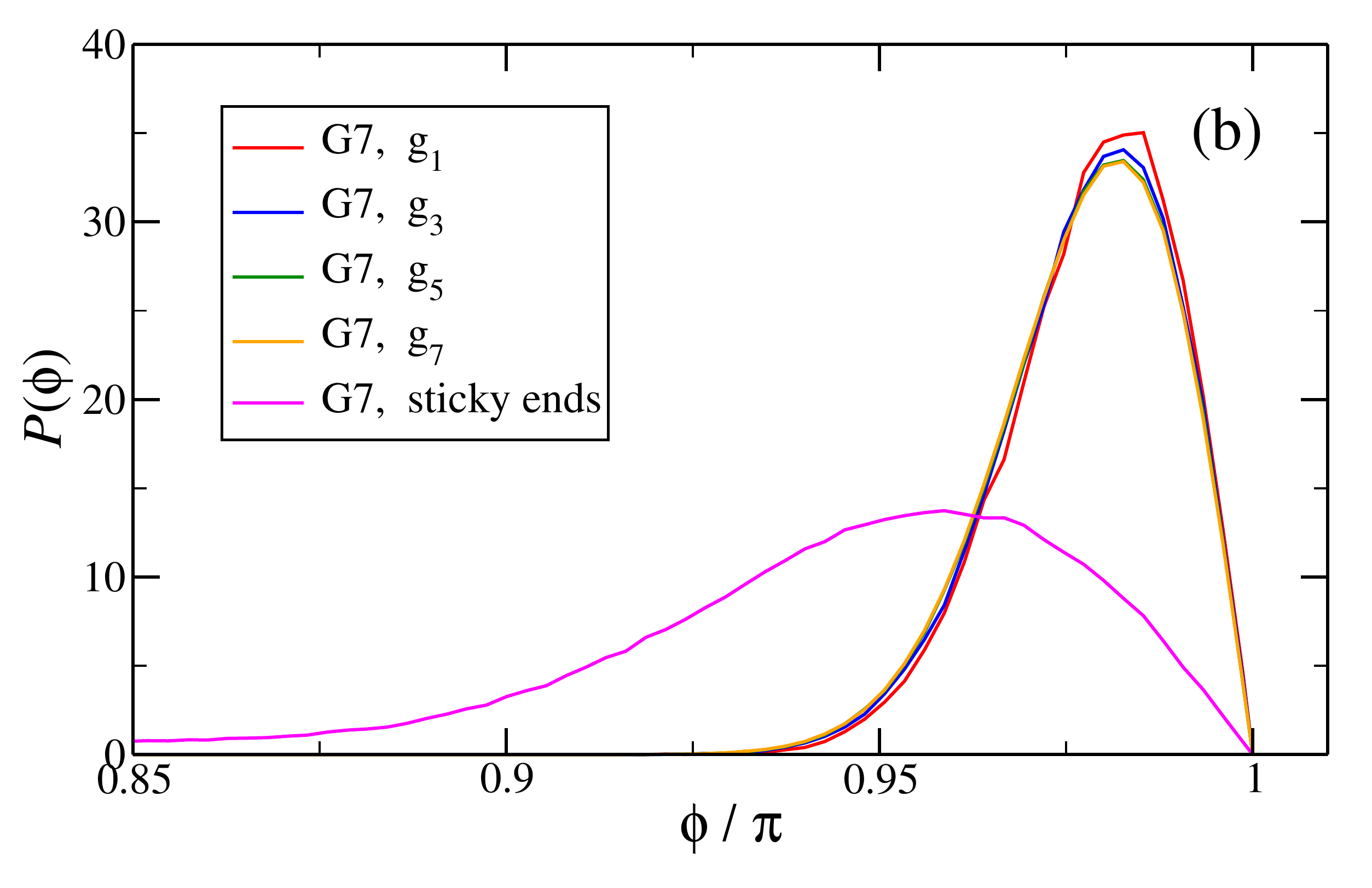}\end{subfigure}}
    \caption{
      (color online) The probability distribution \(P(\phi)\) of angle \(\phi\) of (a) the innermost subgeneration \(\mathrm{g_1}\) for DL-DNAs of generation index \(\mathrm{G1}\)-\(\mathrm{G7}\) (as labeled) and (b) individual subgenerations within a \(\mathrm{G7}\) DL-DNA (as labeled). The distribution
      \(P(\phi)\) is normalized as \(\int_0^\pi \! P(\phi) \, \mathrm{d}\phi = 1\).
    }
    \label{fig:fig8}
  \end{center}
\end{figure}

In order to analyze the internal freedom of the typical conformations of the dendrimers, we measure in the simulations two kinds of bond angles, namely \(\phi\) and \(\theta\), for the individual subgenerations. 
Here, \(\phi\) is defined, according to eq.~(\ref{vphi:eq}), as the angle between two consecutive bonds within a Y-arm;
it is consequently a reliable  measure of the rigidity of the Y-DNA arms. 
The angles \(\theta_i\) (\(i = 1, 2, 3\)), on the other hand, are defined as the angles between the vectors pointing along the three arms of the Y-DNA, whereby an arm vector is defined as the vector connecting the 
first and last monomer of a specific Y-DNA arm, i.e., the arm is assumed to be fully rigid. 
Each Y-DNA element is characterized by three of these angles: \(\theta_1\), \(\theta_2\), and \(\theta_3\).
For fully rigid connections between successive Y-junctions in the dendrimer, one would find \(\phi = \pi\) and \(\theta_i = 2\pi/3\) for all \(i\).

In Fig.~\ref{fig:fig8}(a) the probability distribution \(P(\phi)\) of the angle \(\phi\) for the innermost generations, $g_1$, is shown for DL-DNAs of different generation index. 
The distributions all exhibit a pronounced maximum close to the value of a fully rigid dendrimer, i.e., \(\phi \cong \pi\). 
This feature again demonstrates that the interactions of our model tend to keep the monomer chains straight. A slight but visible 
enhancement of the peak (i.e., a reduction of the fluctuations) can be seen for the G7-dendrimer. This indicates, that its inner generation
is more stretched and thus more straight as the number of generation grows. This feature is one manifestation of the increased osmotic 
stretching force from the counterions. If we focus on a G7 molecule and look at the stretching of the various generations 
$g_i$ within the G7 DL-DNA, Fig.~\ref{fig:fig8}(b), a similar trend can be observed: the distribution is rather sharply peaked 
close to the angle $\phi = \pi$, pointing to stretched connections between the junction points.   
Again, one can notice that the probability distribution \(P(\phi)\) for higher subgenerations displays a 'leakage' to smaller \(\phi\)-values, indicating that the branches belonging
to the outermost sub-generations are more flexible compared to the innermost branches, which is consistent with the finding on the inter-bonding separation, (see Fig.~\ref{fig:fig7})
and our interpretation of its physical origin. In addition, we also show in Fig.~\ref{fig:fig8}(b)
the probability distribution of angle $\phi$ corresponding to those parts of the chains that are characterized as a sticky-end. 
As it is expected, the function spreads over a wider range of possible conformational angles, reflecting the fact that sticky ends exhibit significantly more flexible behavior.  

\begin{figure}
  \begin{center}
  \centering{\begin{subfigure}{.5\textwidth}\includegraphics[width=0.6\textwidth, angle=270]{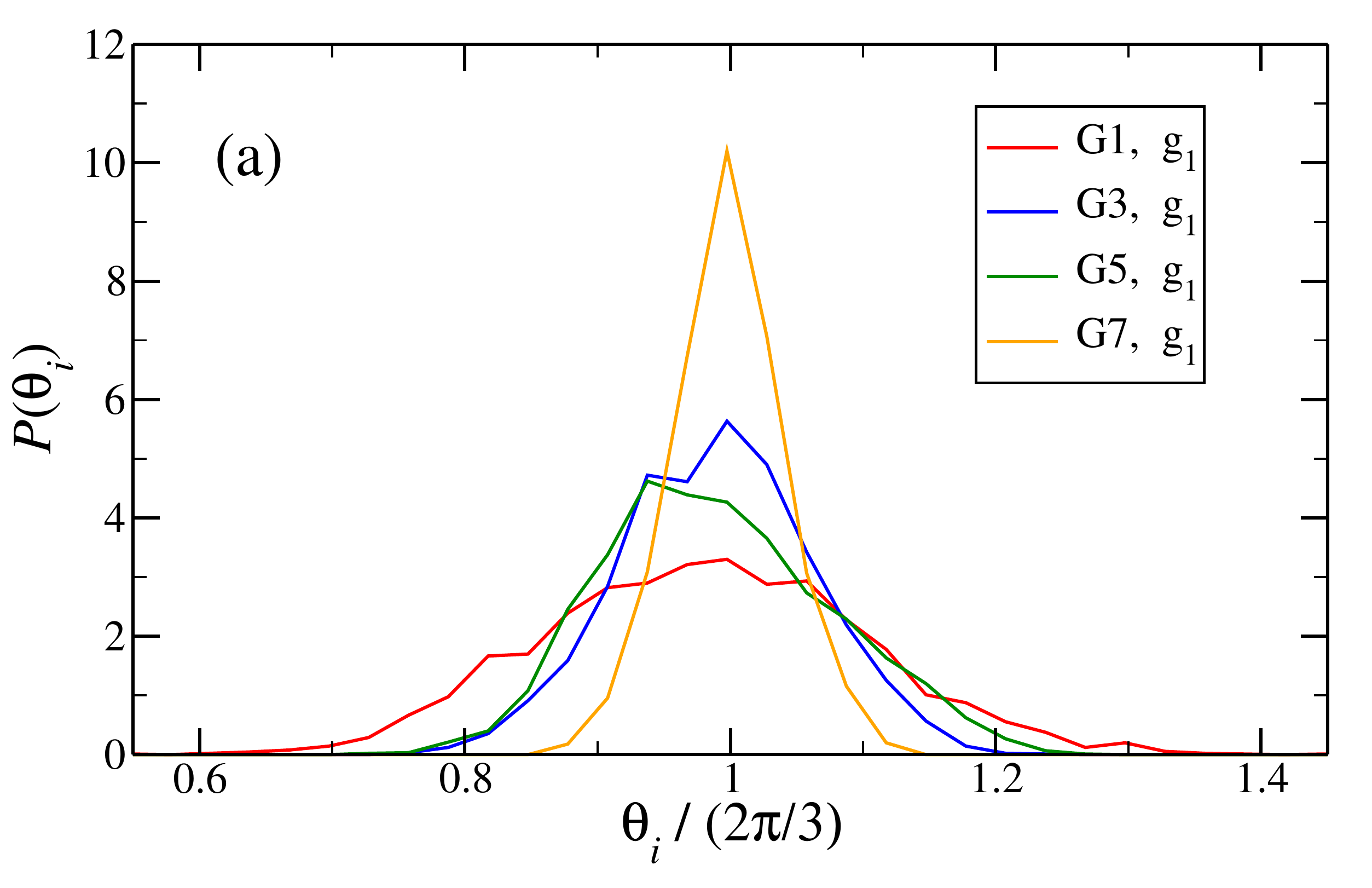}\end{subfigure}\begin{subfigure}{.5\textwidth}\includegraphics[width=0.9\textwidth]{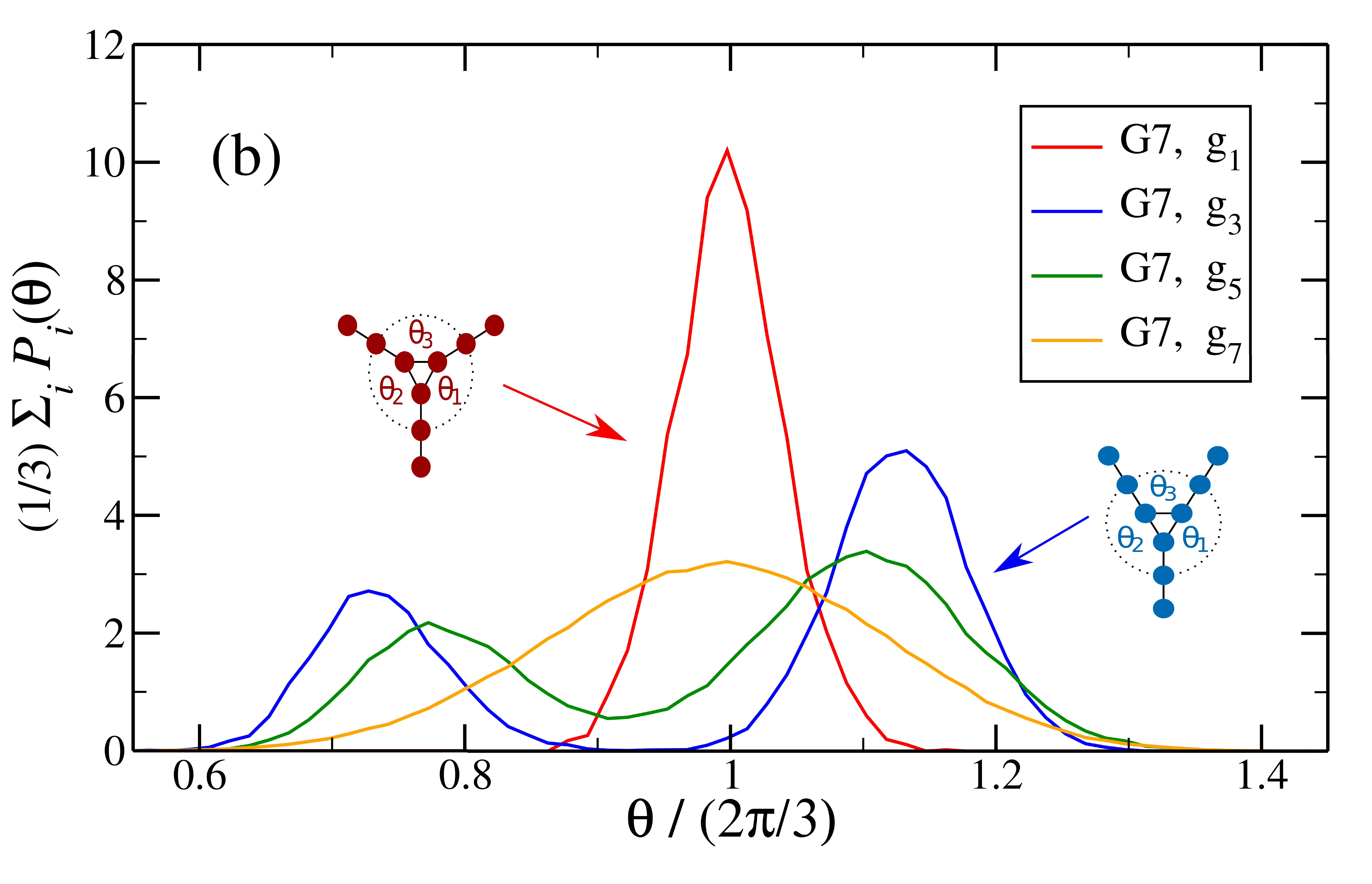}\end{subfigure}}
    \caption{
      (color online) \(P(\theta_i)\), i.e., the probability distribution of angles \(\theta_i\), (\(i = 1, 2, 3\)) of (a) the innermost subgeneration \(\mathrm{g_1}\) for 
      DL-DNA of generation index \(\mathrm{G1}\)-\(\mathrm{G7}\) (as labeled) and (b) the normalized sum over the three probability distributions \(P_i(\theta)\) of angles \(\theta_i\) of individual subgenerations 
      within a \(\mathrm{G7}\) DL-DNA (as labeled).
     Each \(P(\theta_i)\) is normalized as \(\int_0^\pi \! P(\theta_i) \, \mathrm{d}\theta_i = 1\).
    }
    \label{fig:fig9}
  \end{center}
\end{figure}

Fig.~\ref{fig:fig9} shows the corresponding probability distributions \(P(\theta_i)\) for the angles \(\theta_i\), (\(i = 1, 2, 3\)), for DL-DNAs of different generation index. 
In particular, in Fig.~\ref{fig:fig9}(a) we focus on the innermost generation, ${\rm g}_1$, of various G$N$-dendrimers and we do not distinguish between the three angles
$\theta_i$, $i = 1, 2, 3$, since their distributions coincide by symmetry. The distinction is being made in Fig.~\ref{fig:fig9}(b), for which we first collected statistics to determine
the three individual distributions $P_i(\theta)$ for the three angles denoted in the insets, and then we plotted their normalized sum $(1/3)\sum_{i=1}^3 P_i(\theta)$ against $\theta$.
Here, we can see that the most probable angle of the innermost sub-generation, \(\mathrm{g_1}\), of dendrimers of different generations is centered around \(\theta_i = 2\pi/3\), confirming the rigidity of the Y-branches. 
However, the width of these distributions increases with decreasing generation number \(\mathrm{G}N\), as the amplitude of the fluctuations in \(\theta_i\) 
correlates negatively with the size of the dendrimer branch attached 
to the corresponding arm.
With growing generation number, the number of branches grows more rapidly, so that the fluctuations in the angle $\theta$ become less probable due to the reduced available volume and the 
restrictions due to the mutual electrostatic repulsions between the different arms.
When examining the distributions of different sub-generations within a G7 DL-DNA, Fig.~\ref{fig:fig9}(b), the emergence of double peaks at intermediate sub-generations can be observed. 
This phenomenon can be explained by the deformation of Y-DNAs from a conformation with \(\theta_i = 2\pi/3\), \(i = 1, 2, 3\), 
into a configuration with \(\theta_i > 2\pi/3\), with \(i = 1, 2\), and \(\theta_3 < 2\pi/3\).
This change is caused by the monomers pertaining to the outer (inner) generations that pull (push), respectively, monomers of the intermediate subgenerations outwards via steric and electrostatic interactions.
This phenomenon does not occur for Y-DNAs of the last sub-generation, i.e., \(\mathrm{g_7}\) in the examined case, as these Y-DNAs are not constrained by generations of higher index.

\begin{figure}
  \begin{center}
        \includegraphics[width=0.3\textwidth, angle=270]{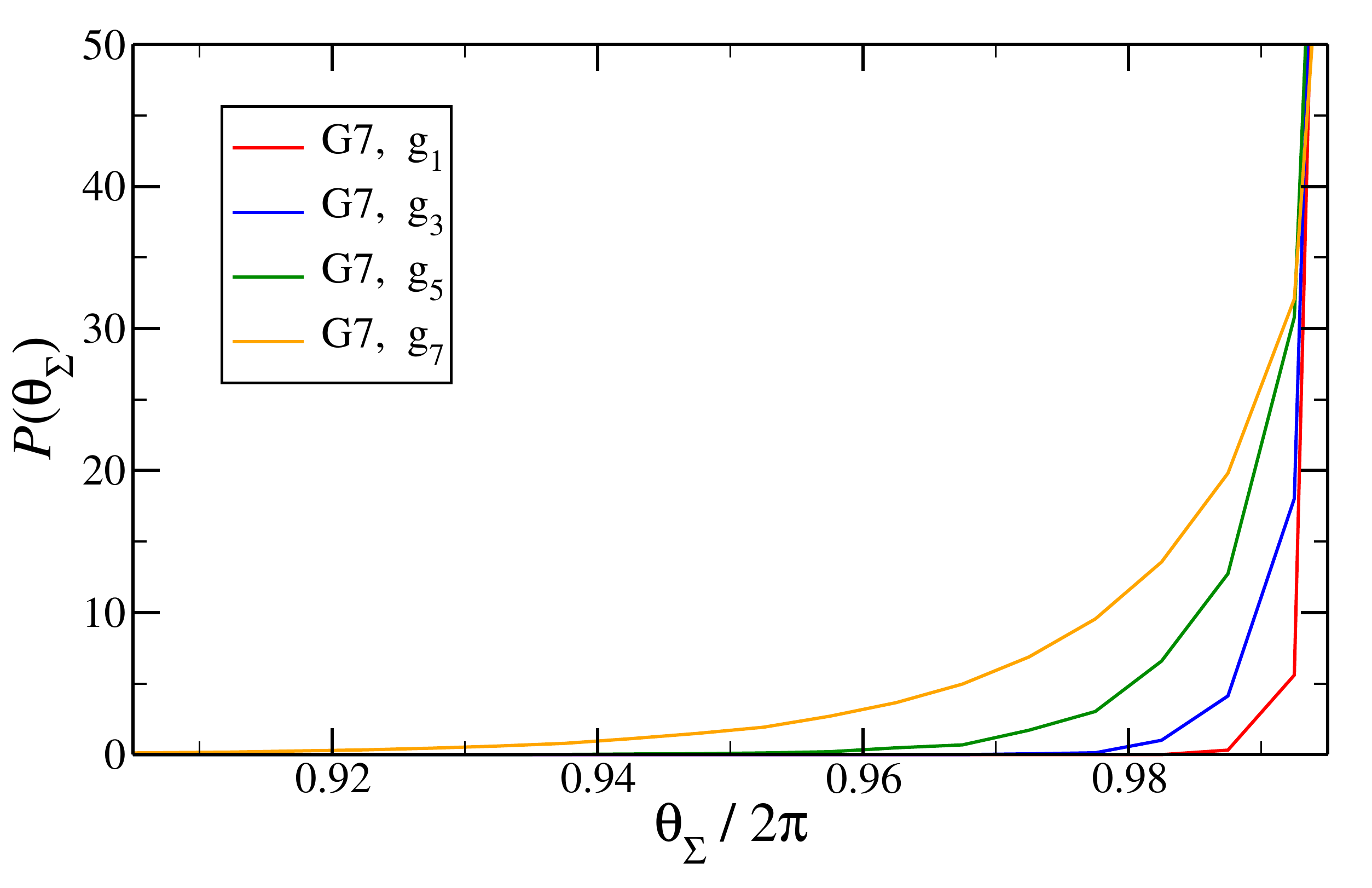}
    \caption{
      (color online) \(P(\theta_\Sigma)\), i.e., the probability distribution of the sum of the three junction angles \(\theta_\Sigma\) of the individual subgenerations within a \(\mathrm{G7}\) DL-DNA (as labeled).
      \(P(\theta_\Sigma)\) is normalized as \(\int_0^{2\pi} \! P(\theta_\Sigma) \, \mathrm{d}\theta_\Sigma = 1\).
    }
    \label{fig:fig10}
  \end{center}
\end{figure}

Finally, in Fig.~\ref{fig:fig10} the distribution \(P(\theta_{\Sigma})\) is shown for a \(\mathrm{G7}\) DL-DNA, where \(\theta_\Sigma = \sum_{i = 1}^3\theta_i\).
The data provide evidence that the Y-DNA of the innnermost subgeneration, whose arms are subject to outward forces caused by the subsequent generations, is almost completely 
planar, i.e., \(\theta_\Sigma \approx 2\pi\).
With increasing generation index, however, the Y-DNAs' deviation from the planar configuration becomes more pronounced, i.e., \(\theta_\Sigma < 2\pi\).
Underlying to this behaviour are two opposite effects: 
Coulomb repulsion and the aforementioned outward forces drive the Y-DNAs towards a planar configuration, but at the same time this planarity reduces the number of configurations available to the Y-DNAs and 
therefore their entropy. As we proceed to the outermost generations, the branches of the Y-junctions have more freedom to fluctuate and entropic contributions become enhanced,
enabling fluctuations of the Y-junctions that deviate from planarity. 

\begin{figure}[h]
  \begin{center}
    \includegraphics[width=0.6\textwidth]{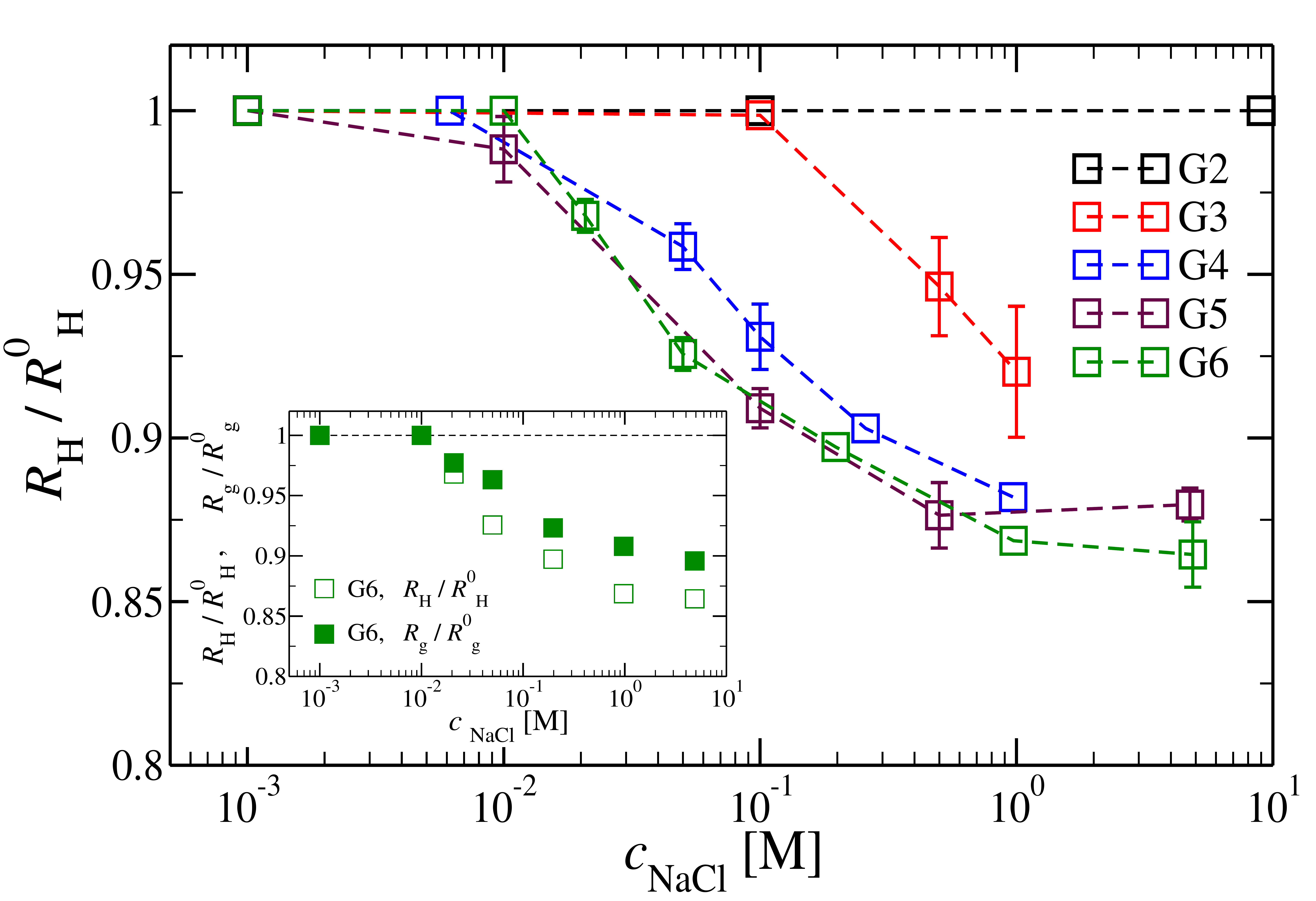}
    \caption{
      (color online) The effect of different NaCl concentrations on the hydrodynamic radius \(R_\mathrm{H}\) for DL-DNA of generations G2 to G6 (as labeled). The inset shows comparison between the $R_{\rm H}$- and the $R_{\rm g}$-shrinkage (as compared to respective salt-free values $R_{\rm H}^0$ and $R_{\rm g}^0$) for a G6 DL-DNA.
    }
    \label{fig:plt_rg_salt_comp}
  \end{center}
\end{figure}

{\bf Influence of salt on conformational properties of DL-DNA.} -- We have also analyzed the effect of a finite salt concentration (NaCl) on the overall size of the dendrimer.
The experiments are performed for different values of salt concentrations, extending from a low salt regime (\(c = \SI{0.1}{\milli\Molar}\)) up to very high salt concentrations (\(c \approx \SI{5}{\Molar}\)).
Due to numerical limitations, only salt concentration up to \(c \approx \SI{30}{\milli\Molar}\) could be considered in simulations. 
Results for the hydrodynamic radius, as obtained from the experiment, are summarized in Fig.~\ref{fig:plt_rg_salt_comp}.
By adding  salt essentially no change in the size of the dendrimer is observable up to a concentration of \(c = \SI{10}{\milli\Molar}\).

\begin{table}[h]
  \begin{center}
  \caption{Comparison of the results for the radius of gyration, \(R_\mathrm{g}\), obtained in simulations and for the hydrodynamic radius, \(R_{\mathrm{H}}\),
 extracted from experiment (as labeled) over six 
generations of DL-DNA. Experiments are performed for a salt concentration of \(\SI{0.1}{\milli\Molar}\). Simulations are carried out both for the  salt-free regime (\(c_0 = \SI{0}{\milli\Molar}\)) and using a salt 
concentrations of \(c_1 = \SI{1}{\milli\Molar}\) with different simulation packages (as labeled).}
    \label{tbl:tab3}
    \begin{tabular}{lllll}
  Generation & \(R_{\mathrm{H/c=\SI{0.1}{\milli\Molar}}}^{\mathrm{exp}}[\si{\nano\meter}]\) & \(R_{\mathrm{g}/c=\SI{0}{\milli\Molar}}^{\mathrm{ESPResSo}}[\si{\nano\meter}]\) & \(R_{\mathrm{g}/c=\SI{0}{\milli\Molar}}^{\mathrm{LAMMPS}}[\si{\nano\meter}]\) & \(R_{\mathrm{g}/c=\SI{1}{\milli\Molar}}^{\mathrm{LAMMPS}}[\si{\nano\meter}]\)\\ \hline
  G1 & 3.5 & 3.25 & 3.37 & --- \\
  G2 & 9.31 & 9.43 & 9.37 &  9.6  \\
  G3 & 14.42 & 15.61 & 15.52 & 15.4  \\
  G4 & 21.43 & 21.68 & 22.02 & 21.6  \\
  G5 & 30.58 & 28.61 & 28.75 & 28.2  \\ 
  G6 & 40.49 & 35.63 & 35.8 & --- \\ 
\end{tabular}
  \end{center}
\end{table}

This observation is confirmed in the simulations, (see Table \ref{tbl:tab3}).
The absence of any significant shrinking at low salt concentrations is the consequence of the rigidity of the molecule, i.e., the high persistence length. 
In order to overcome the stiffness of the molecule, one has to proceed to higher salt concentrations, i.e., above \(c = \SI{10}{\milli\Molar}\); 
under such conditions the screening of the charge of the molecule starts to affect the Coulomb interaction between the monomers, inducing thereby the shrinking of the molecule. 
This reduction in size is more pronounced for higher dendrimer generations and it can range from approximately \(10\%\) to \(15\%\) for extremely high salt concentrations (i.e., \(c \sim \SI{1}{\Molar}\)). 
Throughout, the decrease of \(R_\mathrm{H}\) is generations dependent and the critical salt concentration at which the molecule starts to shrink differs from generation to generation. 
The inset of Fig.~\ref{fig:plt_rg_salt_comp} shows the comparison of the experimental results for the radius of gyration and for the hydrodynamic radius of G6 DL-DNA under the change of salt concentration. The both quantities show the same trend but with slightly more expressed shrinkage of \(R_\mathrm{H}\) over \(R_\mathrm{g}\). 

\begin{figure}[h]
  \begin{center}
 \centering{\begin{subfigure}{.5\textwidth}\includegraphics[width=0.6\textwidth, angle=270]{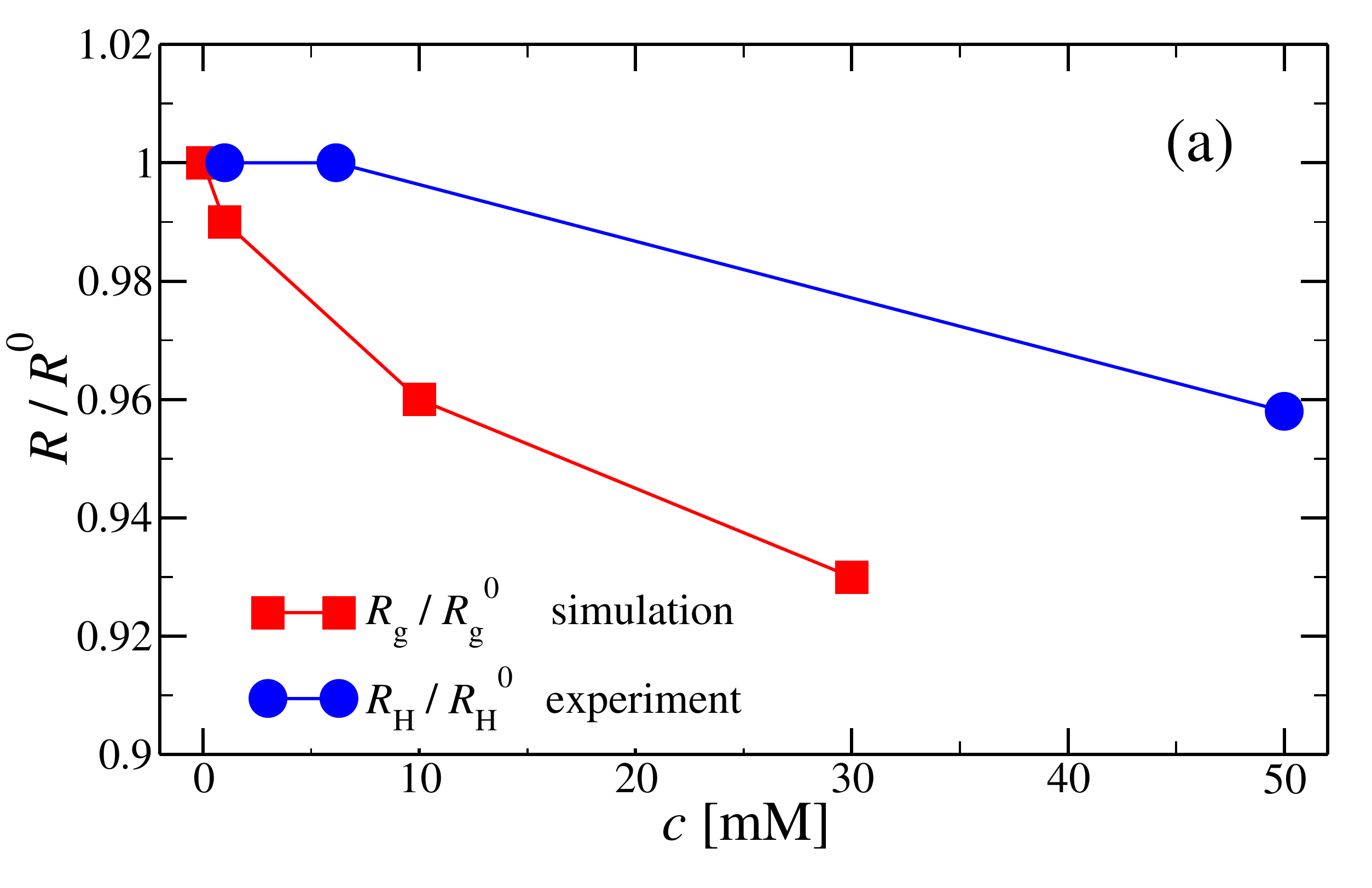}\end{subfigure}\begin{subfigure}{.5\textwidth}\includegraphics[width=0.6\textwidth, angle=270]{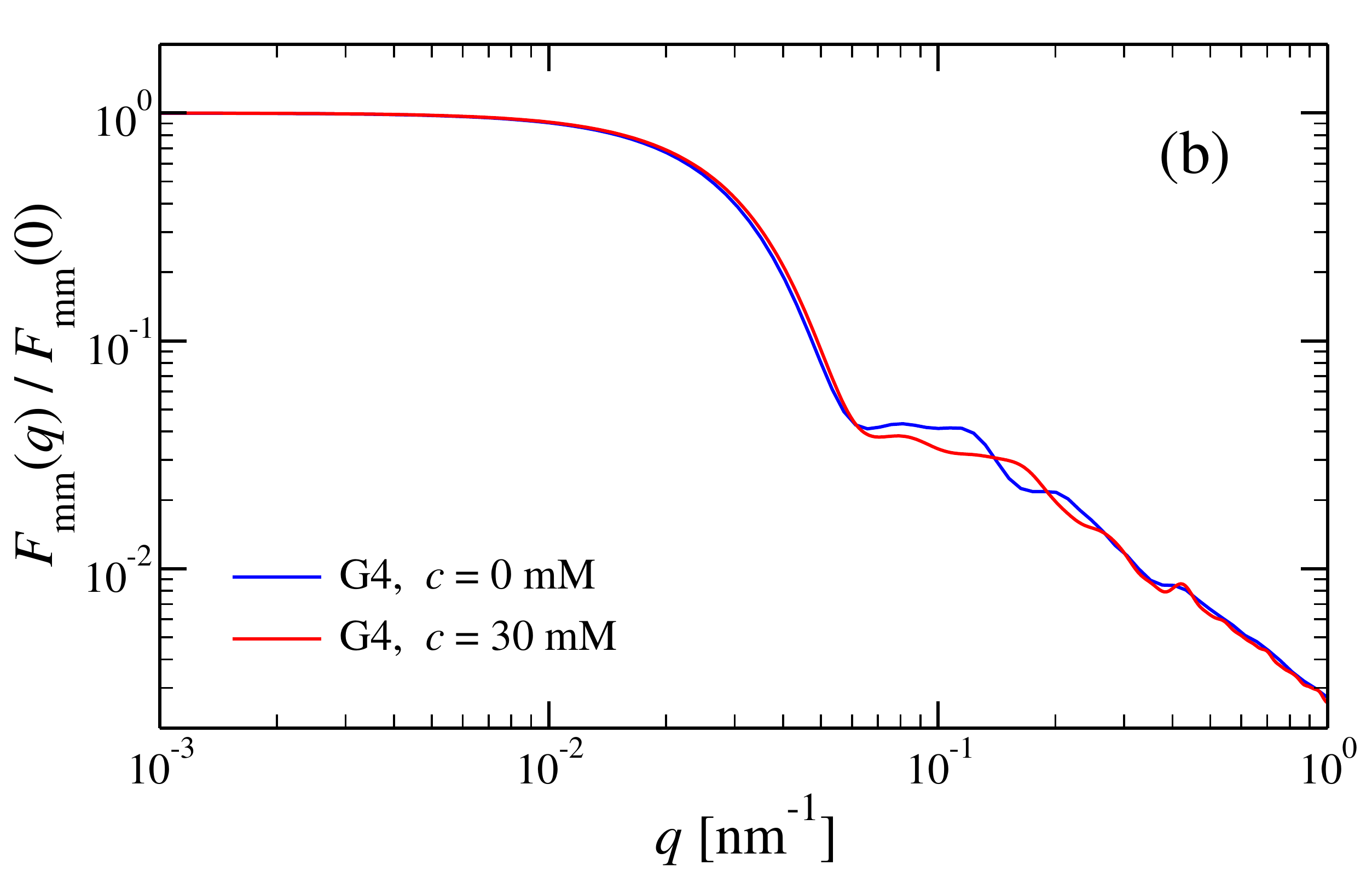}\end{subfigure}}
    \caption{
     (color online) (a) Hydrodynamic radius \(R_\mathrm{H}\), as extracted from experiments, and radius of gyration \(R_\mathrm{g}\) obtained from simulation (as labeled), as functions of the NaCl concentration (given in \si{\milli\Molar}) for a G4 DL-DNA. (b) Form factor \(F_\mathrm{mm}(q)\) as a function of \(q\) for a \(\mathrm{G4}\) DL-DNA for salt concentrations \(c = \SI{0}{\milli\Molar}\) and \(c = \SI{30}{\milli\Molar}\) (as labeled), obtained from simulation.
  }
    \label{fig:plt_rg_salt_g4}
  \end{center}
\end{figure}

\begin{figure}
\centering{\begin{subfigure}{.5\textwidth}\includegraphics[width=0.6\textwidth, angle=270]{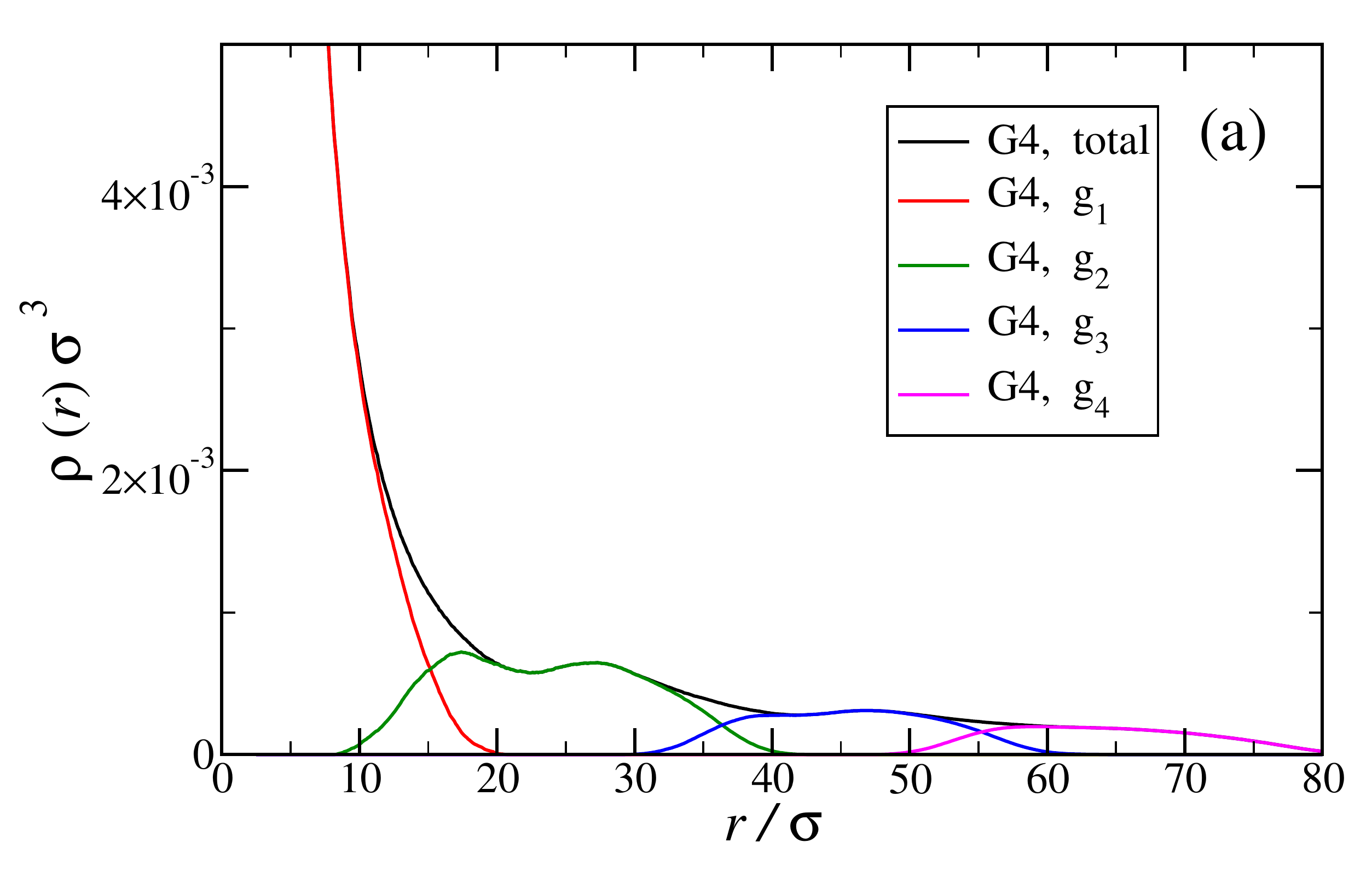}\end{subfigure}\begin{subfigure}{.487\textwidth}\includegraphics[width=0.62\textwidth, angle=270]{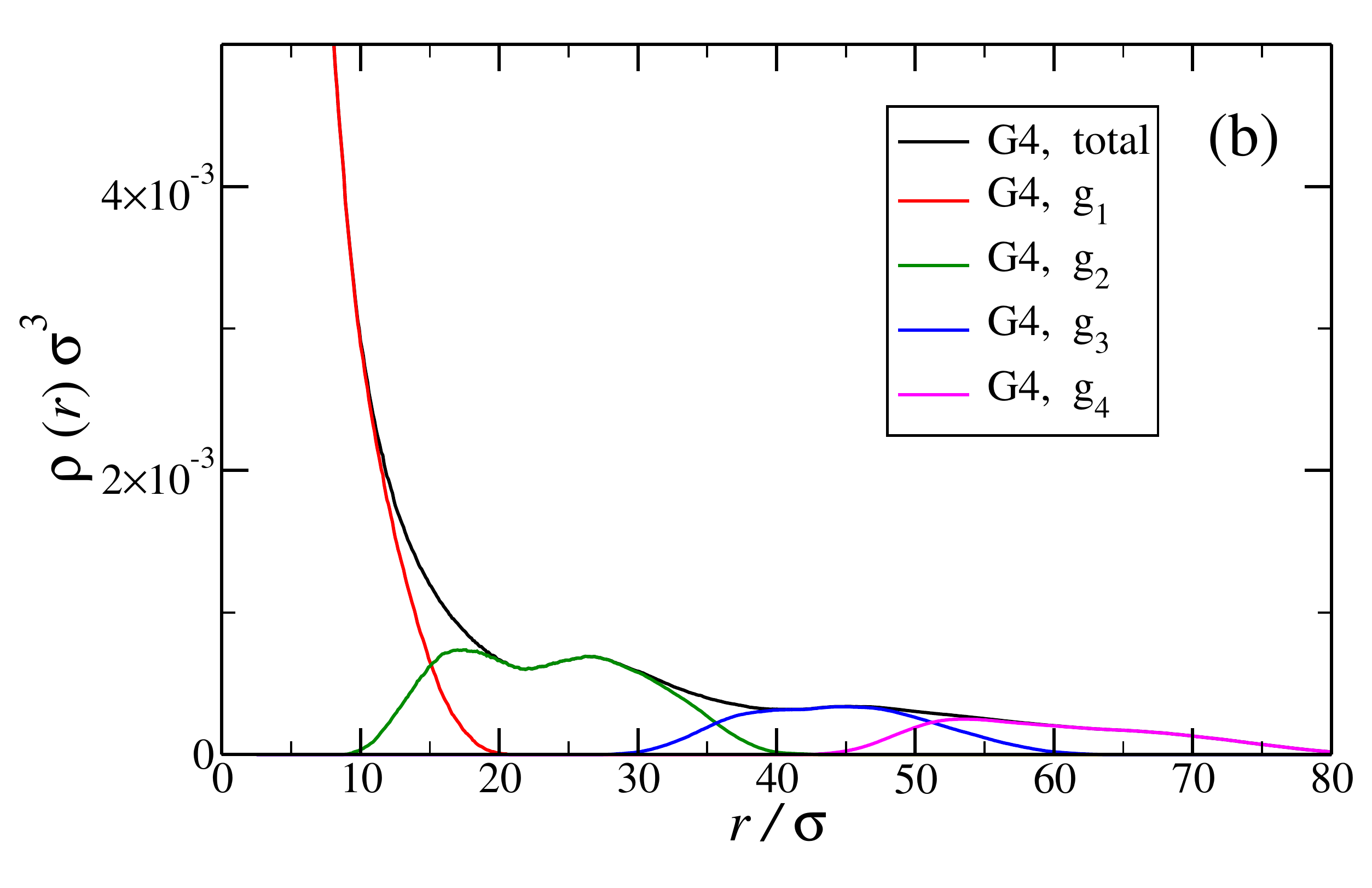}\end{subfigure}}
\caption{Monomer density profile for G4 DL-DNA: (a) salt-free regime, $c = 0$ mM; (b) $c = 30$~mM.  }
\label{fig:G4ro}
\end{figure}

We have performed simulations of G4 DL-DNA for various values of salt concentration, and radius of gyration as a function of salt concentration is given in Fig.~\ref{fig:plt_rg_salt_g4}(a). The comparison with the hydrodynamic radius obtained from experiments is also shown. 
From simulation, we observe the shrinking of the molecule by approximately \(7\%\) varying the salt concentration from \(c=\SI{1}{\milli\Molar}\) to \(c=\SI{30}{\milli\Molar}\), while the experiments show the shrinkage by approximately \(5\%\) varying the salt concentration from \(c=\SI{1}{\milli\Molar}\) to \(c=\SI{50}{\milli\Molar}\).
In Fig.~\ref{fig:plt_rg_salt_g4} (b) we present the form factors, comparing two sets of simulation data corresponding to different salt regimes. 
In correspondence with the above discussed results for \(R_\mathrm{H}\) and \(R_\mathrm{g}\), a small shrinkage is observable in the case of a finite salt concentration 
reflected by a somewhat slower decay of \(F_\mathrm{mm}(q)\) as a function of \(q\) for \(q \in [10^{-2},  10^{-1}]\) $\rm nm^{-1}$. 
Nevertheless, the curves unambiguously demonstrate that the added salt does not have any impact on the rigidity of the dendrimer, i.e., in the regime of large wave vectors the form factor satisfies the above 
mentioned $q^{-1}$ scaling law, irrespective of the salt concentration.
From these data, one can conclude that the Coulomb interactions do stretch the bonds between the monomers, but the related effect is subdominant as compared to the role that rigidity has in suppressing 
significant changes in bond lengths.

The shrinking of flexible, charged polymers upon addition of salt can be physically traced back to the screening of the electrostatic interaction or to the enhanced
osmotic pressure from the co- and counterions at the exterior of the macromolecule~\cite{jusufi:jcp:02}. For the case at hand, it is not clear that the same physical mechanism
is at work since the shrinkage is minimal and the molecular architecture is different. To shed light into the mechanism behind the size reduction of the nanoparticle,
we look at the density profiles with and without salt.
In Fig.~\ref{fig:G4ro}(a), the monomer density profile for G4 DL-DNA is shown for salt-free case and for $\rm NaCl$ salt solution of concentration $c = 30$ mM 
in Fig.~\ref{fig:G4ro}(b). One can notice that the only 
difference between corresponding monomer profiles under different salt conditions appears for the outermost sub-generation $g_4$. In the salt solution, enhanced backfolding of the outermost branches 
arrises, resulting in the small shrinkage of dendrimer's radius of gyration, observed at Fig.~\ref{fig:plt_rg_salt_g4}(a). The interior of the dendrimer remains unaffected by the
added salt, an additional manifestation of the combined effect of rigidity and branched architecture of the novel, DL-DNA constructs. These are remarkably resistant macromolecules,
which nevertheless feature a very low internal monomer concentration, allowing them to absorb counterions or smaller molecules in their interior. In addition, whereas salinity
is expected to affect the effective interactions between such dendrimers, practically it does not affect their sizes and shapes, rendering them thus as prime candidates for molecules
with tunable, ultrasoft effective repulsions.

\section{Conclusions}

We have investigated the structural properties of DL-DNA at a single particle level with sizes ranging from G1 to G7. 
Additionally, we probed the salt-responsiveness of the complex nanostructure of said molecules with regards to backfolding of dendritic arms,
providing a thorough investigation of the structural properties of these all-DNA nanostructures.
Through a combination of experiments and molecular modeling, we provide 
an advancement of our understanding of such dendritic DNA constructs, which is essential for developing applications and investigating novel phenomena 
related to this type of soft material. 

In colloidal polymer network terms, the high-G DL-DNA and ionic microgels  share some common characteristics. Both, highly permeable to solvent molecules, can act as efficient absorber of their own counterions under salt-free conditions, and in very coarse-grained level their internal structure has a core-shell morphology. However, our results revealed that the DL-DNA's scaffold architecture and its inherent rigidity grand these all-DNA nanotructures with low internal monomer concentration, regular voids in their interior and, at the same time, a resilience against the addition of salt; intriguing and promising features which are absent in ionic microgels and they are expected to have significant impact on dendrimers' collective behavior. More specific, experimental as well as computational results show that varying the salt concentration only minimally affects the conformation and does not cause any backfolding of dendritic arms. This low salt-responsiveness allows for adjusting the effective interaction between different DL-DNA molecules without the dendrimers collapsing or their structure deforming significantly. In addition, the monomer density profiles revealed that high-G DL-DNA are dendrimers with almost flat-density and internal cavity with generation-independent size. The cavity space, located at the dendrimer's center of mass, found to be comparable to the size of G1 implying that the void interior can be engineered at subnanometer precision (at the level of a single level) by simply adjusting the arm length of the Y-DNA building block belonging to the first generation; thus allowing full control over the degree of dendrimer interpenetrability. We believe that the  above properties, make these highly charged tailored empty-core/shell nanostructures ideal candidates for exploring novel forms of self-assembly such as cluster crystallization in the bulk. The current work set the pivotal point for the investigation on the many-body properties of concentrated DL-DNA systems, which will be the subject of the future work.


\section{Methods}

Here, we briefly specify some of the experimental as well as simulation techniques employed in this work.

{\bf DNA sequences and synthesis of DL-DNA.} -- The DL-DNA nanostructures were fabricated following the synthetic procedure described in reference~\cite{NatProtoc2006}, which is based on two assembly 
procedures: (a) self-assembly and (b) enzyme-assisted assembly. Briefly, DL-DNA was prepared from a core three-arm DNA junction (Y-DNA), having each arm terminated with a non-palindromic four-base-long 
sticky-end. We designate this Y-DNA as a first dendrimer generation (G1). To build up the next generation (G2), the above all-DNA tri-functional core was hybridized with three other Y-DNAs with sticky-ends 
complementary to the core Y-DNA. The cohesions points were enzymatically sealed using T4 DNA ligase (Promega). Additional generations
(G3, G4, G5, etc.) were created by repeatedly ligating Y-DNAs to the
sticky-ends of the previous generation. The Y-DNA building block is synthesized by annealing of three partially complementary single-stranded DNAs (ssDNAs) at equal molar ratio,  employing a 
one-pot approach. 

The sequences of DNA strands used to create the DL-DNA constructs are slightly modified compared to those reported in reference~\cite{NatProtoc2006}, in order to minimize the total number of different strands 
necessary for synthesizing all-DNA dendrimers up to the $6^{\rm th}$ generation. The sequences of DNA strands were designed using the program SEQUIN~\cite{SEQUIN}. All DNA strands used in this study 
were purchased from Integrated DNA Technologies, Inc. (www.idtdna.com), phosphorylated at their 5'-end and purified by denaturing polyacrylamide gel electrophoresis. The DNA strand concentrations were 
determined by measuring the absorbance at $260$ nm with a micro-volume spectrometer (NanoDrop 2000).

DNA sequences and construction scheme for synthesizing the all-DNA dendrimers used in this study, are listed below. The bold letters correspond to sticky-end sequence and p indicates the position of the 
phosphate modification.

\vskip 0.2in

 
$\Box$ $Y_{1a}$ : 5'-p-\textbf{TGAC} TGGATCCGCATGACATTCGCCGTAAG-3'

$*$ $Y_{2a}$ : 5'-p-\textbf{GTCA} TGGATCCGCATGACATTCGCCGTAAG-3'

$\Diamond$ $Y_{3a}$ : 5'-p-\textbf{ATCG} TGGATCCGCATGACATTCGCCGTAAG-3'

$\triangle$ $Y_{4a}$ : 5'-p-\textbf{GCAA} TGGATCCGCATGACATTCGCCGTAAG-3'

$\Box$ $Y_{1b}$ : 5'-p-\textbf{TGAC} CTTACGGCGAATGACCGAATCAGCCT-3'

$*$ $Y_{2b}$ : 5'-p-\textbf{CGAT} CTTACGGCGAATGACCGAATCAGCCT-3'

$\Diamond$ $Y_{3b}$ : 5'-p-\textbf{TTGC} CTTACGGCGAATGACCGAATCAGCCT-3'

$\triangle$ $Y_{4b}$ : 5'-p-\textbf{GTCA} CTTACGGCGAATGACCGAATCAGCCT-3'

$\Box$ $Y_{1c}$ : 5'-p-\textbf{TGAC} AGGCTGATTCGGTTCATGCGGATCCA-3'

$*$ $Y_{2c}$ : 5'-p-\textbf{CGAT} AGGCTGATTCGGTTCATGCGGATCCA-3'

$\Diamond$  $Y_{3c}$ : 5'-p-\textbf{TTGC} AGGCTGATTCGGTTCATGCGGATCCA-3'

$\triangle$ $Y_{4c}$ : 5'-p-\textbf{GTCA} AGGCTGATTCGGTTCATGCGGATCCA-3'

\vskip 0.2in


$1^{st}$ generation DL-DNA (G1): G1 = $Y_{1}$ = $Y_{1a} + Y_{1b} + Y_{1c}$

$2^{nd}$ generation DL-DNA (G2): G2 = G1 + $3\times Y_{2}$, where $Y_{2} = Y_{2a} + Y_{2b} + Y_{2c}$

$3^{rd}$ generation DL-DNA (G3): G3 = G2 + $6\times Y_{3}$, where $Y_{3} = Y_{3a} + Y_{3b} + Y_{3c}$

$4^{th}$ generation DL-DNA (G4): G4 = G3 + $12\times Y_{4}$, where $Y_{4} = Y_{4a} + Y_{4b} + Y_{4c}$

$5^{th}$ generation DL-DNA (G5): G5 = G4 + $24\times Y_{5}$, where $Y_{5} = Y_{1a} + Y_{2b} + Y_{2c}$

$6^{th}$ generation DL-DNA (G6): G6 = G5 + $48\times Y_{6}$, where $Y_{6} = Y_{3} = Y_{3a} + Y_{3b} + Y_{3c}$

\vskip 0.2in

\begin{figure}[h]
  \begin{center}
  \includegraphics[width=0.4\textwidth]{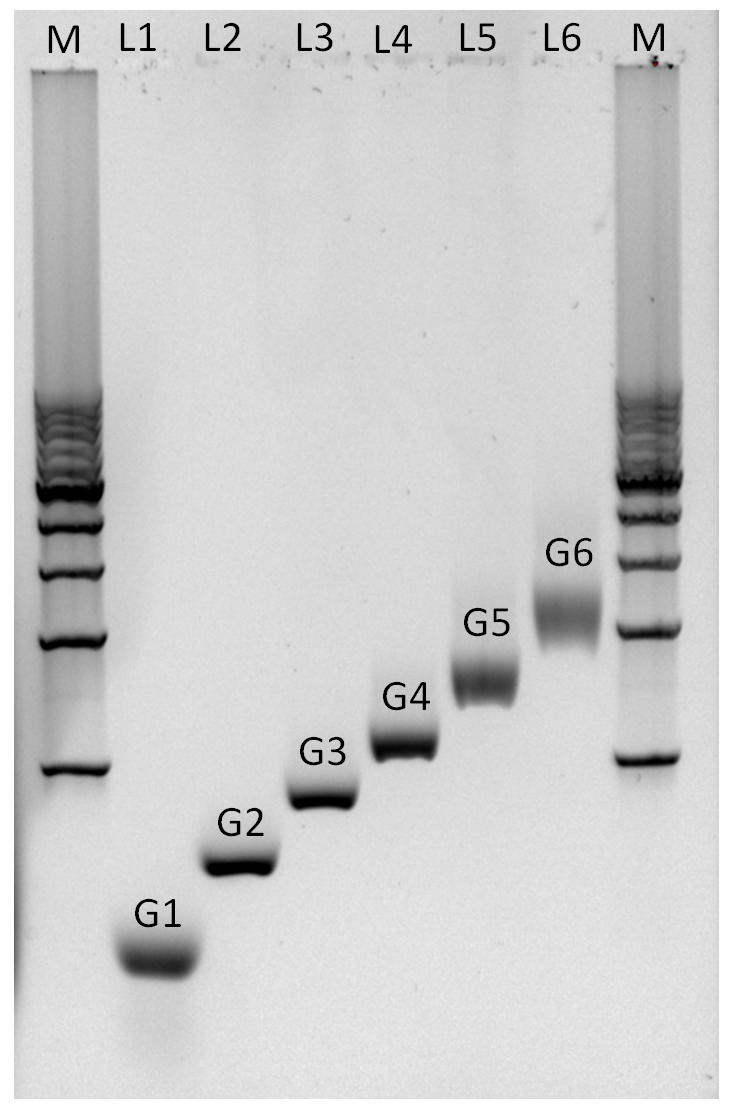}
    \caption{Non-denaturing agarose gel ($0.5\%$) electrophoresis analysis of the assembly of DL-DNA. The electrophoretic mobility of all-DNA dendrimers up to 6th generation is demonstrated. Line M: 1 Kbp 
    double-stranded DNA marker (bottom to the top: 1 Kbp to 10 Kbp with a 1 Kbp step), Line 1: G1, Line 2: G2, Line 3: G3, Line 4: G4, Line 5: G5 and Line 6: G6.}
    \label{fig:electrophoresis}
  \end{center}
\end{figure}

Agarose gel electrophoresis was employed to confirm the successfull assembly of all-DNA dendrimers. As shown in Figure \ref{fig:electrophoresis}, the desired DL-DNA constructs migrates as single sharp bands, 
showing decreasing mobility with increasing generation.

\begin{figure}[h]
  \begin{center}
  \includegraphics[width=0.75\textwidth]{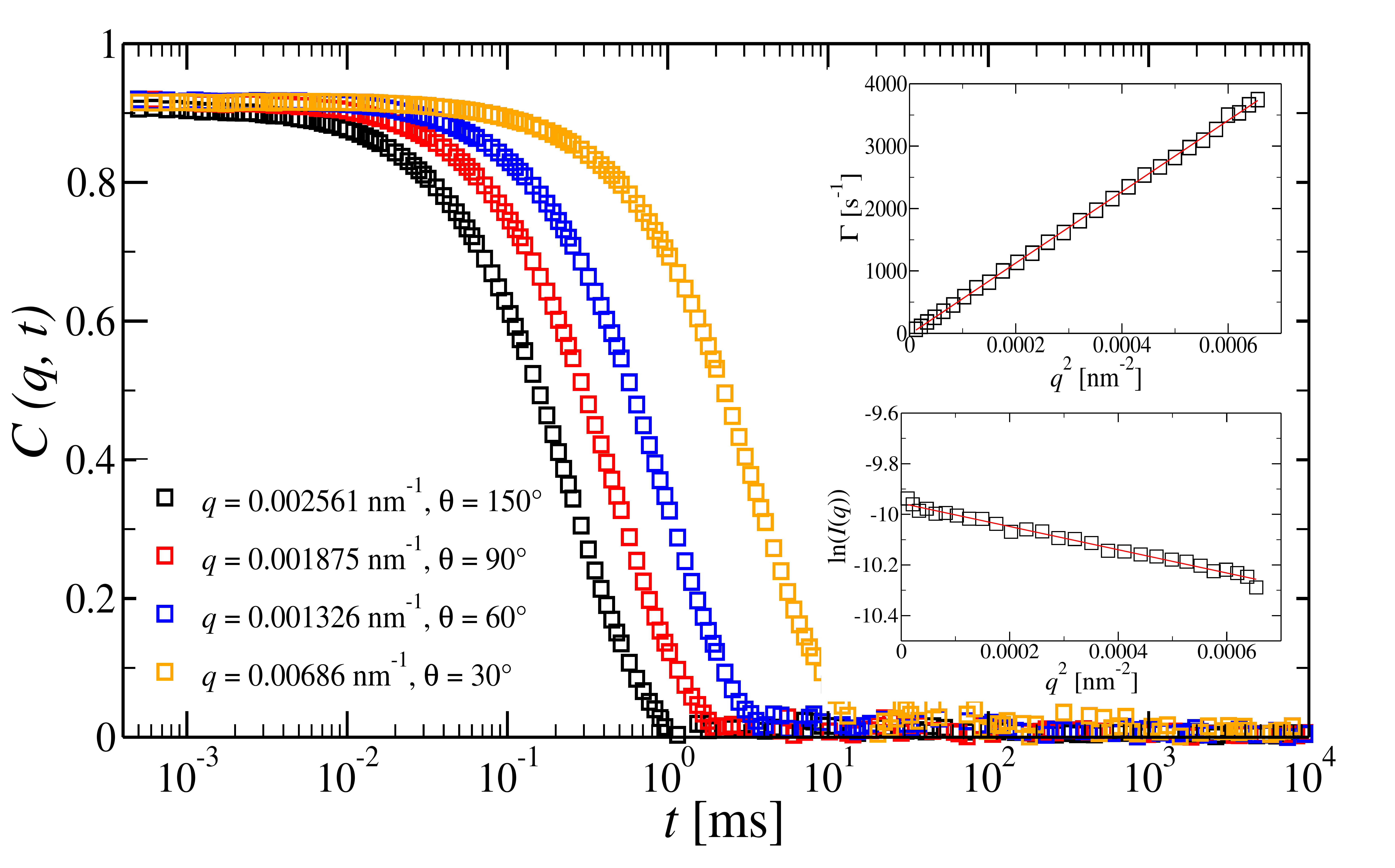}
    \caption{Intermediate scattering functions, $C(q,t)$, at different angles for a dilute G6
all-DNA dendrimer aqueous solution at a concentration of $\SI{15}{\nano\Molar}$  and under mild-salinity
conditions ($\SI{50}{\milli\Molar}$  NaCl). For such dilute regime only a single translational diffusion
process is observed. Indeed, the $C(q,t)$ is a single diffusive relaxation which is clearly
confirmed by the $q^2$ dependence of the decay rate, $D=\Gamma$ (top inset). Bottom inset: The
natural logarithm of the reduced static light scattering intensity for the G6 dendrimer as a
function of $q^2$ (Guinier plot). In both insets, the slopes (red lines) were used for the
extraction of the $R_{\rm H}$ (top) and $R_{\rm g}$ (bottom).}
    \label{fig:G6cR}
  \end{center}
\end{figure}

\vskip 0.2in
{\bf Light scattering experiments.} -- Dynamic light scattering (DLS) and static light scattering (SLS) experiments were performed by employing an ALV goniometer setup equipped with an Helium-Neon laser 
operating at $\lambda=632.8$ nm. The effective hydrodynamic radii of the DL-DNA constructs in dilute aqueous solutions at different conditions of salinity (NaCl) were measured with DLS. The Brownian motion of 
the DL-DNA molecules was recorded in terms of the time auto-correlation function of the polarized light scattering intensity $G(q,t)$, using an  ALV-5000  multi-tau  digital correlator. The measurement consisted of 
obtaining the intermediate scattering (field) function $C(q,t)=\sqrt{(G(q,t)-1)/\beta}$ at several scattering wave vectors $q=(4{\pi}n/\lambda)(\sin{(\theta)}/2)$, where $\beta$ is an instrumental factor related to the 
spatial coherence constant and depends only on the detection optics, $n$ the refractive index of the solvent and $\theta$ the scattering angle. $C(q,t)$ was analyzed via an Inverse Laplace Transform (ILT) using 
the CONTIN algorithm~\cite{DLS} and the average relaxation time was determined from the peak of the distribution of relaxation times. The translational diffusion coefficient, $D=\Gamma/q^2$, was found to be 
$q$-independent ($\Gamma$ is the measured relaxation rate). The hydrodynamic radius was extracted from the measured diffusion coefficient $D$ assuming validity of the Stokes-Einstein relation, 
$R_{\rm H}=kT/6{\pi\eta}D$ ($k$ is the Boltzmann constant, $T$ the absolute temperature and $\eta$ the solvent viscosity) for spherical objects, (see Fig \ref{fig:G6cR}). 

The radius of gyration $R_{\rm g}$ of the higher DL-DNA generations was extracted from SLS
experiments in very dilute aqueous solutions. The scattered intensity of the solution $I_{\rm sol}$, the
solvent  $I_{\rm solv}$ and the pure toluene  $I_{\rm tol}$ were recorded over an angular range from $15^o$ to $150^o$ corresponding to scattering vector q in range of 
$3.46\times 10^3 < q < 2.55\times 10^2\ \rm nm^{-1}$. The pure
toluene was used as a scattering-angle-independent standard to account the dependence of
the scattering volume on the scattering angle. Thus, the scattered intensity $I$ of the DLDNA
particles was determined as follows: $I(q) =(I_{\rm sol}(q)-I_{\rm solv}(q))/I_{\rm tol}(q)$. The $R_{\rm g}$ was
obtained from SLS experiments by performing a Guinier plot: $\ln{(I(q))} = \ln{(I(0))}-(q^2R_{\rm g}^2)/3$. All experiments reported here were performed at room temperature.

{\bf Simulation description.} -- To simulate the behaviour of the above described macromolecular system with given interactions we employ molecular dynamic (MD) simulations.
Simulations are performed using two independent platforms, namely ESPResSo~\cite{espresso1,espresso2} and LAMMPS~\cite{lammps}. 
The time steps used is \(\Delta t = 10^{-3}\tau\), where $\tau = \sqrt{m\sigma^2/\epsilon}$, so that the total running time of the simulations extends over \SI{10}{\nano\second} (i.e., over $10^8$ simulation steps). 
The box size is chosen to keep the total particle density, $\rho_{\rm p}$, independent of dendrimer generation. 
This number density is set to the value of \(\rho_{\rm p} = \SI{5e-7}{\angstrom^{-3}}\) in order to avoid self-interaction via periodic boundaries while keeping the box size small at the same time.

Canonical ensemble is applied and periodic boundary condition are applied. 
The evaluation of electrostatic interactions is performed using the Ewald summation method~\cite{ewald} and the multilevel summation method~\cite{msm1,msm2} (MSM) for ESPResSo and LAMMPS, 
respectively, with a relative force accuracy of \(10^{-5}\). 
Langevin thermostat is chosen and set to preserve the temperature of \(T = \SI{298}{\kelvin}\).

\begin{acknowledgement}
This work has been supported by the Austrian Science Fund (FWF) under Grant number I 2866-N36
and by the Deutsche Forschungsgemeinschaft (DFG) under Grant number STI 664/3-1.
Computation time at the Vienna Scientific Cluster (VSC) is gratefully acknowledged.
\end{acknowledgement}

\providecommand{\latin}[1]{#1}
\providecommand*\mcitethebibliography{\thebibliography}
\csname @ifundefined\endcsname{endmcitethebibliography}
  {\let\endmcitethebibliography\endthebibliography}{}




\end{document}